\documentclass[aps,prl,amsmath,amssymb,twocolumn]{revtex4-2}

\usepackage{graphicx}
\usepackage{dcolumn}
\usepackage{footmisc}

\usepackage{bm}
\usepackage{hyperref}
\usepackage{xcolor}
\usepackage{float}

\begin{document}
\setlength{\abovedisplayskip}{5pt}
\setlength{\belowdisplayskip}{5pt}
\setlength{\abovedisplayshortskip}{5pt}
\setlength{\belowdisplayshortskip}{5pt}

\preprint{}

\title{Tidal Disruption Events and Dark Matter Scatterings with Neutrinos and Photons}

\author{Motoko Fujiwara\textsuperscript{1}, Gonzalo Herrera\textsuperscript{2,1,3}}
\affiliation{\textsuperscript{1}Physik-Department, Technische Universit\"at M\"unchen, James-Franck-Stra\ss{}e, 85748 Garching, Germany,\\ \textsuperscript{2} Center for Neutrino Physics, Department of Physics, Virginia Tech, Blacksburg, VA 24061, USA,\\ \textsuperscript{3} Max-Planck-Institut f\"ur Physik (Werner-Heisenberg-Institut), F\"ohringer Ring 6,80805 M\"unchen, Germany,}

\begin{abstract}
	Stars can be tidally disrupted when passing near a black hole, and the debris can induce a flux of high-energy neutrinos. It has been discussed that there are hints in IceCube data of high-energy neutrinos produced in Tidal Disruption Events. The emitting region of neutrinos and photons in these astrophysical events is likely to be located in the vicinity of the central black hole, where the dark matter density might be significantly larger than in the outer regions of the galaxy. We explore the potential attenuation of the emitted neutrino and photon fluxes due to interactions with dark matter particles around the supermassive black hole of the host galaxies of AT2019dsg, AT2019fdr and AT2019aalc, and study the implications for some well-motivated models of dark matter-neutrino and dark matter-photon interactions. Furthermore, we discuss the complementarity of our constraints with values of the dark matter-neutrino scattering cross section proven to alleviate some cosmological tensions.
\end{abstract}

\maketitle

\section{Introduction}
\label{sec:Intro}

\begin{figure}[t]
		\centering
		\includegraphics[width=0.49\textwidth]{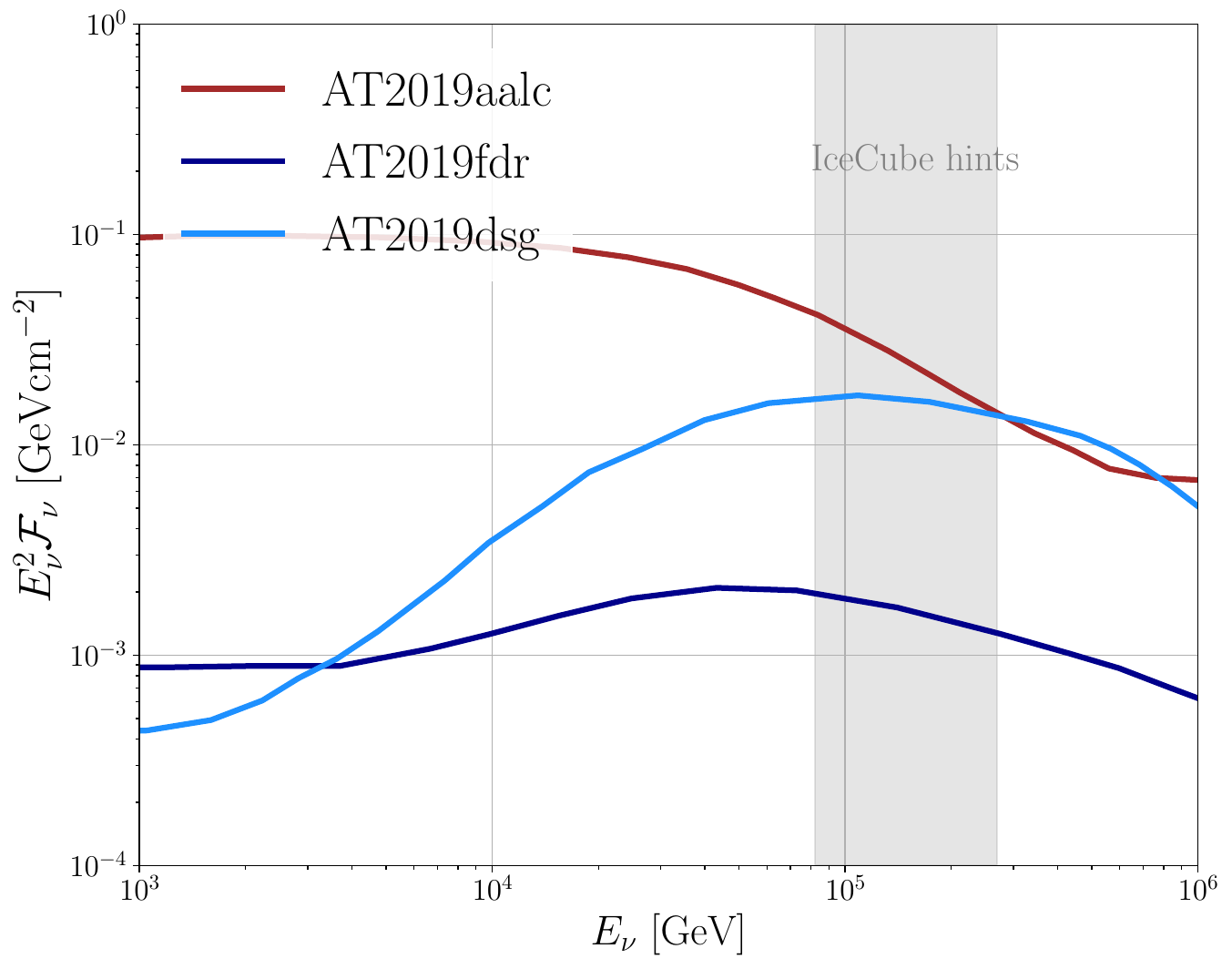}
		\centering
		\caption{Time-integrated neutrino fluxes (fluences) expected from the TDEs AT2019aalc, AT2019fdr and AT2019dsg, in a model with moderate energy protons predominantly scattering on the X-ray target at the source, from \cite{Winter:2022fpf}. For comparison, we show in grey the energy range where IceCube may have detected events from these sources.}
		\label{fig:fluences}
	\end{figure}

Tidal Disruption Events (TDEs) occur when a star passes within a critical distance from a black hole, and is disrupted by the strong tidal force. The accretion of the released stellar matter in the vicinity of the black hole produces flares of photons in several wavelengths lasting for several days~\cite{1975Natur.254..295H, Rees:1988bf, Komossa:2015qya}. It has been discussed that TDEs may be sources of ultra high energy cosmic rays, which would inevitably lead to the production of high-energy neutrinos \cite{2008AIPC.1065..201M, Wang_2011, Farrar:2014yla, Pfeffer:2015idq, Dai:2016gtz,Lunardini:2016xwi,Senno:2016bso,Biehl:2017hnb, Zhang:2017hom, Guepin:2017abw,Murase:2020lnu, Chan:2021blg, Mukhopadhyay:2023mld, Yuan:2023cmd}. In recent years, independent groups have claimed the detection of three TDEs in IceCube data with a 3.7 $\sigma$ combined significance \cite{Stein:2020xhk, Reusch:2021ztx, vanVelzen:2021zsm, Winter:2022fpf}. The TDEs associated with high energy neutrinos are labeled as AT2019dsg, AT2019fdr and AT2019aalc, with neutrino energies of 270 TeV, 82 TeV and 176 TeV, respectively. Further, it has been discussed that the hints of neutrino fluxes from these TDEs are compatible with the acceleration and production mechanisms present at these sources, under certain conditions~\cite{Lunardini:2016xwi,Winter:2022fpf} (See Fig \ref{fig:fluences}).

The photon and neutrino flares from TDEs are crucial to understand the matter distribution in the innermost regions of the galaxy and the physical processes happening in the vicinity of supermassive black holes (SMBH). Moreover, TDEs are especially important since they are typically associated with SMBHs that are quiet and therefore more difficult to study than Active Galactic Nuclei (AGN), since they show no steady emission. In particular, it has recently been discussed that the neutrino and photon observations from some AGN (TXS 0506+056 and NGC 1068 \cite{IceCube:2018dnn, IceCube:2022der}) can be used to probe interactions between Dark Matter (DM) and the Standard Model (SM) particles in the vicinity of the central black hole \cite{Ferrer:2022kei, Cline:2022qld, Cline:2023tkp, IceCube:2023cwx, Herrera:2023nww}. 

This novel phenomenological probe allows to constrain smaller values of the DM-neutrino and DM-photon scattering cross section than complementary constraints in certain regions of the parameter space, \textit{E.g} \cite{Choi:2019ixb, Kelly:2018tyg, Alvey_2019, Arg_elles_2017, Murase:2019xqi,Brax:2023tvn,Carpio:2022sml,A:2023wup,Lin:2023nsm,Koren:2019wwi,McMullen:2021ikf,Mangano_2006,Wilkinson:2013kia, Wilkinson:2014ksa, Escudero:2015yka}, and in certain regions of the parameter space of DM-proton and DM-electron interactions, \textit{E.g} \cite{Essig:2017kqs,XENON:2019gfn,Jho:2021rmn,Zhang:2020nis,Ghosh:2021vkt,Farzan:2014gza,Das:2021lcr, Lin:2022dbl, Wang:2021jic,Super-Kamiokande:2022ncz, Alvey:2022pad, An:2021qdl, Guo:2020oum, Ambrosone:2022mvk, Bell:2023sdq, John:2023knt}.

\begin{table*}[t!]
  \renewcommand{\arraystretch}{1.5}
		\begin{center}
			\begin{tabular}{c||cc|ccc|c}
				\hline
				& $M_{\rm BH}$ & $\Delta M_{\rm BH}$ & $R_{\rm em}$ & $\gamma_{\rm sp}$ & $\langle\sigma v \rangle$/$m_{\rm DM}$ & $\rho_{\rm DM}(R_{\rm em})$  \\
				\hline
				\toprule
				 AT2019dsg (1)
                  & 
                  & 
                  & ~$1.6 \times 10^{-4}$ pc~
                  & ~2.25~ 
                  & 0 
                  & $5.2\times10^{15}$ GeV/cm$^{3}$   
                  \\
				 AT2019dsg (2)
                  & 5 $\times 10^{6}M_{\odot}$ 
                  & 0.5 $\times 10^{6}M_{\odot}$ 
                  & $1.6 \times 10^{-4}$ pc 
                  & 1.5 
                  & 0 
                  & $2.1\times10^{13}$ GeV/cm$^{3}$      
                  \\
				 AT2019dsg (3) 
                  & 
                  & 
                  & 0.016 pc 
                  & 1.5 
                  & ~$10^{-30}$cm$^{3}$s$^{-1}$/GeV~
                  & ~$2.2\times10^{10}$ GeV/cm$^{3}$~
                  \\
                  \hline
				 AT2019fdr (1) 
                  & 
                  & 
                  & 0.0015 pc 
                  & 2.25 
                  & 0 
                  & $3.6\times10^{13}$ GeV/cm$^{3}$  
                  \\
				 AT2019fdr (2) 
                  & ~1.3 $\times 10^{7}M_{\odot}$~
                  & ~0.13 $\times 10^{7}M_{\odot}$~
                  & 0.0015 pc 
                  & 1.5 
                  & 0 
                  & $4.3\times10^{11}$ GeV/cm$^{3}$   
                  \\
				 AT2019fdr (3) 
                  & 
                  & 
                  & 0.81 pc 
                  & 1.5 
                  & $10^{-30}$cm$^{3}$s$^{-1}$/GeV 
                  & $6.1\times10^{7}$ GeV/cm$^{3}$   
                  \\
                  \hline
				 AT2019aalc (1) 
                  & 
                  & 
                  & 0.0016 pc 
                  & 2.25 
                  & 0 
                  & $3.8\times10^{13}$ GeV/cm$^{3}$  
                  \\
				 AT2019aalc (2) 
                  & 1.6 $\times 10^{7}M_{\odot}$ 
                  & 0.16 $\times 10^{7}M_{\odot}$ 
                  & 0.0016 pc 
                  & 1.5 
                  & 0 
                  & $ 3.7 \times10^{11}$ GeV/cm$^{3}$   
                  \\
                  AT2019aalc (3) 
                  & 
                  & 
                  & 0.065 pc 
                  & 1.5 
                  & $10^{-30}$cm$^{3}$s$^{-1}$/GeV 
                  & $ 1.5\times10^{9}$ GeV/cm$^{3}$   
                  \\
                \hline
			\end{tabular}
		\end{center}
		\caption{Relevant parameters for the TDE candidates considered in this work, for three different sets of assumptions dubbed (1), (2), and (3). 
        $M_{\rm BH}$ is the mass of the central black hole, $\Delta M_{\rm BH}$ the uncertainty on the measurement of the mass of the central black hole, $R_{\rm em}$ is the region where neutrino and photon are emitted, $\langle\sigma v \rangle$/$m_{\rm DM}$ denotes the assumed values of the effective DM self-annihilation cross section, and $\rho_{\rm DM}(R_{\rm em})$ is the density of DM particles in the region where the TDE occurs.}
		\label{tab:TDEs}
	\end{table*}
Observations of photons and hints of neutrinos suggest that the emitting region of TDEs is likely to be located very close to the SMBH, at $R_{\rm  em} \sim 10^2-10^3 R_{\rm  S}$, where the density of DM particles is expected to be very high, potentially forming a spike \cite{Gondolo_1999}. Indeed, the theoretically-expected tidal disruption radius $R_{\rm TDE}$ of a star with mass $M_{*}$ and radius $R_{*}$ can be approximated as \cite{1975Natur.254..295H}
\begin{equation}
R_{\rm TDE} \sim R^*\left(\frac{M_{\rm B H}}{M^*}\right)^{\frac{1}{3}},
\end{equation}
such that, for example, for a star with mass and radius of the Sun, the disruption occurs at a distance from Sagittarius A$^{*}$ of $R_{\rm TDE} \sim 3.7 \times 10^{-6}$ pc $\sim 10 R_{\rm  S}$. Similar values of the tidal radius are expected in general for main sequence stars, which satisfy $R_{*} \sim M_{*}^{0.8}$ \cite{Ryu:2020cqv}. 

It is believed that the neutrino and gamma-ray emission from blazars such as TXS 0506+056 happens farther from the central black hole, a the Broad Line Region, or beyond $R_{\rm em} \sim 10^4 R_{\rm  S}$ \cite{Padovani_2019}. In non-jetted galaxies such as NGC 1068, current observations suggest acceleration and neutrino emission in the corona $R_{\rm em} \sim 10-10^2 R_{\rm  S}$ \cite{Murase:2022dog, Eichmann:2022lxh}, however, uncertainties are large, and the acceleration mechanism of protons to high-energies is still unclear. In this sense, TDE observations may constitute a complementary and perhaps more robust probe of the DM distribution in the inner regions of galaxies, since stars are tidally disrupted only at short distances from the black hole. Further, since TDEs are associated with quiet galaxies, the astrophysical background of photon and neutrino emission is lower, and the DM spike may have been less depleted due to interactions with stars or disrupting processes~\cite{Merritt:2003qk, Bertone:2004pz, Merritt:2006mt}. Therefore, it becomes important to derive constraints on DM-neutrino and DM-photon scatterings from these sources, and confront them with the constraints obtained from other AGNs such as TXS 0506+056.

We notice that the identification of AT2019fdr and AT2019aalc  as TDEs, which occurred in AGNs, is still in debate. Other interpretations include Hydrogen-rich superluminous supernovae~\cite{Pitik_2022}. Although the physics of TDEs is not fully established, we attempt to exploit these astrophysical events to explore the possibility of probing DM scattering properties, using the three TDE neutrino emitting candidates. 

In this work, we first review TDEs and astrophysical models to describe the phenomena quantitatively. In this part, we focus on how we can estimate emission radii of neutrinos, which is crucial for determining sensitivities on DM effects. Second, we calculate the DM distribution in the vicinity of AT2019dsg, AT2019fdr and AT2019aalc, under different assumptions. We then derive upper limits on the DM-neutrino and DM-photon scattering cross sections when assuming that the scattering cross section are constant, comparing our results with complementary constraints. Furthermore, we derive limits in concrete models of DM-neutrino and DM-photon interactions. Finally, we present our conclusions.

\section{Brief review of TDEs}

In this section, we briefly review the observational status and theoretical models of TDEs and specify the important parameters in our analysis. TDEs are caused by the deformation of massive stars due to the tidal force when they pass close to the SMBH~\cite{1975Natur.254..295H,Rees:1988bf}~(see \cite{Komossa:2015qya} for an observational review).  Part of the star is bound to the SMBH to fully accrete, which realizes a month- to a year-long flare. This results in X-ray, optical ultraviolet (OUV), and infrared (IR) observations. Thanks to these observational inputs, the physical picture of TDEs in the post accretion phase have been addressed by models with accretion disk, semi-relativistic outflow, and jet. See \textit{E.g.} Ref.~\cite{Dai_2018}.

Multimessenger astronomy using neutrino signatures is another crucial direction to reveal the physics of TDEs. 
Currently, three events (AT2019~\cite{Stein:2020xhk}, AT2019fdr~\cite{Reusch:2021ztx}, and AT2019aalc) are associated with neutrino events at IceCube (IC191001A, IC200530A, and IC191119A, respectively). Interestingly, the observed neutrino events delay $\sim  {\cal  O}(100)$~days compared to the peak detection in the blackbody photon ({\textit E.g.} Table 1 in~\cite{Winter:2022fpf}). The neutrino production mechanism and the emission region are not established and various models have been proposed~\cite{PhysRevD.84.081301,PhysRevD.93.083005,Wang:2011ip,Lunardini:2016xwi,Senno:2016bso,Hayasaki:2019kjy,Fang:2020bkm,Farrar:2014yla,Zhang:2017hom,Biehl:2017hnb,Guepin:2017abw,Liu:2020isi,Winter:2020ptf,Wu:2021vaw,Murase:2020lnu}, which often focus on a specific TDE. In Ref.~\cite{Winter:2022fpf}, the authors proposed a unified and time-depending model to describe observed three events quantitatively for both photon and neutrino observations. In this model, the observed neutrino delay is related to the physical size of the accretion system, and caused by the confinement of protons and particle propagation. In particular, the models were constructed based on the following common features observed in all three events; (i) observation of X-ray signals, (ii) relatively heavy SMBH masses ($\sim  10^{6.5}$--$10^{7.5} M_\odot$), and (iii) observation of IR dust echo. To reproduce the observed high-energy neutrino, the non-thermal isotropic and time-depending protons injection is considered, where protons should be accelerated to interact with background photons or matter. In this scenario, there exists a threshold for the proton's maximum energy to be consistent with the observed photon energy and high-energy neutrino flux. In particular, photon temperature differs depending on the region, and we can test the validity of each model by comparing the observed flux and theoretical prediction. The three models differ on the energies of the ambient photon targets for protons (X-ray, OUV, and IR photons) and the model features are summarized in Fig.~2, 5, and 8 in~Ref.~\cite{Winter:2022fpf}. Using these models, different possibilities for the neutrino emission radius are systematically studied, which gives us a hint on the neutrino emission region. In each model, the emission radius is characterized by different mechanisms. For example, on the X-ray photon target, the proton is confined due to the magnetic field. Assuming the diffusion coefficient has Bohm scaling and identifying the diffusion time scale as the observed neutrino delay time ($t_{\rm  delay}  \sim  {\cal  O}(100)~\mathrm{days}$), the confined radius is characterized as follows~\cite{Winter:2022fpf}. 
\begin{align}
  R  
  &\sim  \sqrt{\frac{E_p}{e  B}  \cdot  t_{\rm  delay}}
  \nonumber
  \\
  &\sim  0.0016~\mathrm{pc}
  \left( \frac{B}{1~\mathrm{G}} \right)^{-\frac{1}{2}}
  \left( \frac{E_p}{5~\mathrm{PeV}} \right)^\frac{1}{2}
  \left( \frac{t_{\rm  delay}}{500~\mathrm{days}} \right)^\frac{1}{2},
\end{align}
where $B$ denotes the magnetic field, and we take the required maximum proton energy in $p  \gamma$ process with X-ray photon, $E_p  \sim  5~\mathrm{PeV}$. Once we assume the same values for the magnetic field and required proton energy for all three TDEs, the evaluated emission radii are universal. For the OUV photon target, the radius is identified as blackbody radii determined from 
TDE observations (\cite{vanVelzen2021} (see Fig.~6) for AT2019dsg, and~\cite{PhysRevLett.128.221101} for AT2019fdr)
and estimated value for AT12019aalc~\cite{Winter:2022fpf}. The estimated radius may differ depending on TDEs, and the smallest values are obtained for neutrino emission radii as $R_{\rm  em}  \sim  1.6  \times  10^{-4}~\mathrm{pc}$  for AT2019dsg ($R_{\rm  em}  \sim  0.0015~\mathrm{pc}$ for AT2019fdr). For the IR photon target, the highest values are required for maximum proton energy since photons have lower energy compared to that of the X-ray and OUV targets. 
The radius is identified as dust radius, which is taken from radio observation for AT2019dsg ($R_{\rm  em}  \sim  0.016~\mathrm{pc}$, see Fig.~3 of~\cite{Stein:2020xhk}), 
IR observation for AT2019fdr ($R_{\rm  em}  \sim  0.81~\mathrm{pc}$~\cite{PhysRevLett.128.221101}), 
and 
estimated using their dust model for AT2019aalc ($R_{\rm  em}  \sim  0.065~\mathrm{pc}$~\cite{Winter:2022fpf}). 
All the above models can predict consistent diffuse neutrino flux, while the authors conclude that it is premature to determine the neutrino production model from the currently available information. Besides, it has been pointed out that models for AT2019aalc have larger uncertainties in the estimated emission radius compared to those for other TDEs. 
Being aware of these uncertainties, we decided to take the uncertainty range of $R_{\rm  em}$ by combining all  these models (target with X-ray, OUV, and IR photons) for each TDE as summarized below.
\begin{align}
  R_{\rm  em}  &=  [ 1.6  \times  10^{-4},  0.016]~\mathrm{pc},  &&\text{(AT2019dsg)}
  \\
  R_{\rm  em}  &=  [ 0.0015,  0.81]~\mathrm{pc},  &&\text{(AT2019fdr)}
  \\
  R_{\rm  em}  &=  [ 0.0016,  0.065]~\mathrm{pc}.  &&\text{(AT2019aalc)}
\end{align}
Based on the described microscopic pictures, the associated neutrino flux can be estimated. In Fig.~\ref{fig:fluences}, we showed the time-integrated neutrino energy fluxes predicted from the model with X-ray target~\cite{Winter:2022fpf}, where neutrino and photon are expected to be emitted from the same radius.

In the next section, we discuss the attenuation effect via DM and neutrino/photon scattering. In this analysis, the choice of neutrino emission radius turns out to be crucial for the sensitivity to constrain DM because the DM density may differ by orders of magnitude with radii. See the last column in Table~\ref{tab:TDEs} and the discussion below. Besides, the energy spectrum of neutrino flux can be an essential input when deriving constraints on the DM-neutrino energy-dependent cross section. If the cross  section is independent of energy, the attenuation effects can be factorized, and the derived constraint is independent of the initial flux. On the other hand, we need to specify the emitted flux before attenuation to derive the constraint on DM scattering to include the energy transport effect (see the second term of Eq.~\eqref{eq:cascade}). 
As it will be discussed later, we found this effect is only important for the DM-neutrino scattering and use the energy flux of the model with the X-ray target~\cite{Winter:2022fpf}.

\section{DM density in the vicinity of TDE candidates}
\label{sec:DMspike}

Adiabatically-growing black holes form a spike of DM particles in their vicinity \cite{1972GReGr...3...63P,Quinlan:1994ed,Gondolo_1999, Ullio:2001fb}. An initial DM profile of the form $\rho (r) = \rho_0 (r/r_0)^{-\gamma}$ evolves into:
\begin{align}
	\rho_{\rm sp}(r) = \rho_{R} \, g_{\gamma}(r)\, \bigg(\frac{R_{\rm sp}}{r}\bigg)^{\gamma_{\rm sp}}\;,
\end{align}
where $R_{\rm sp}=\alpha_{\gamma}r_0  \left( M_{\rm BH}/(\rho_{0}r_{0}^{3}) \right)^{\frac{1}{3-\gamma}}$ is the size of the spike, with $\alpha_\gamma\simeq 0.293\gamma^{4/9}$ for $\gamma \ll 1$, and numerical values for other values of $\gamma$ provided in \cite{Gondolo_1999}. $\gamma_{\rm sp}=\frac{9-2\gamma}{4-\gamma}$ is the spike slope. Further, $g_{\gamma}(r)$ is a function which  can be approximated for $0<\gamma <2 $ by  $g_{\gamma}(r) \simeq (1-\frac{4R_{\rm S}}{r})^3$, with $R_{\rm S}$ the Schwarzschild radius, while $\rho_{\rm R}$ normalizes the profile, $\rho_R=\rho_{0}\, (R_{\rm 
sp}/r_0)^{-\gamma}$. Our density profile vanishes at $4 R_{\rm S}$, which is a conservative approximation that neglects relativistic and rotating effects in black holes \cite{Sadeghian_2013,Ferrer:2017xwm}.

	\begin{figure}[tb]
		\centering
		\includegraphics[width=0.49\textwidth]{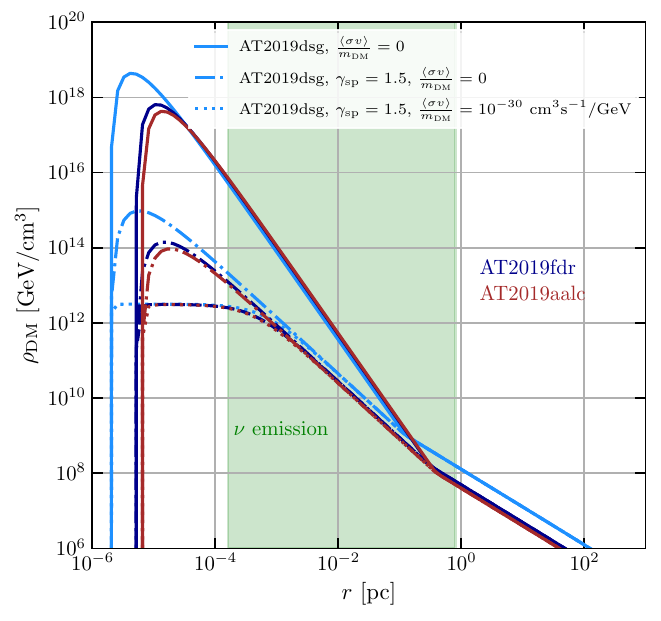}
		\centering
		\caption{DM distribution around the black holes of AT2019dsg, AT2019fdr, and AT2019aalc (light blue, dark blue, and brown), for different values of the DM self-annihilation cross section over its mass. The green region shows the likely emitting region of neutrinos from these sources.}
		\label{fig:DMspike}
	\end{figure}
 
	\begin{figure*}[t!]
		\centering
		\includegraphics[width=0.49\textwidth]{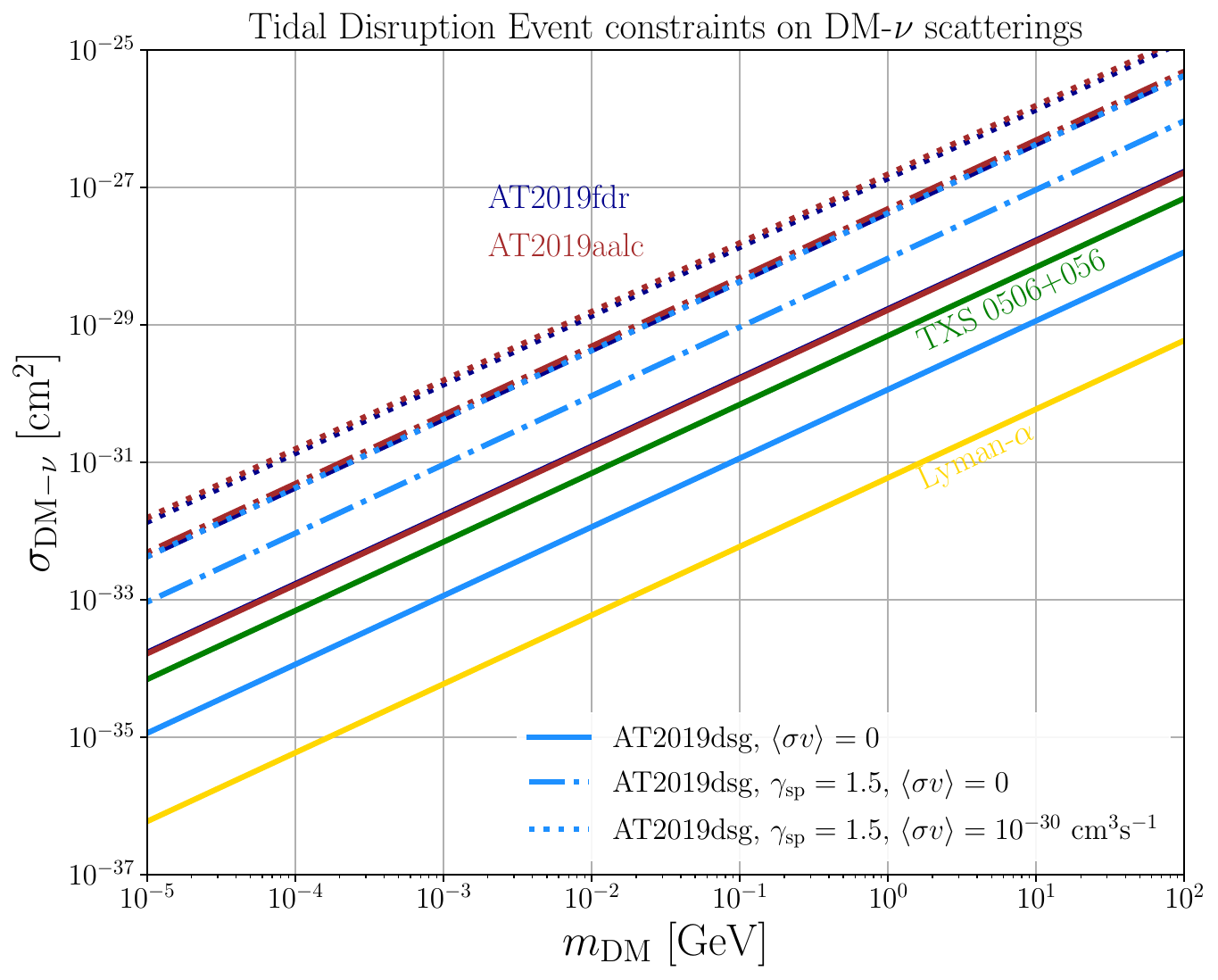}
		\includegraphics[width=0.49\textwidth]{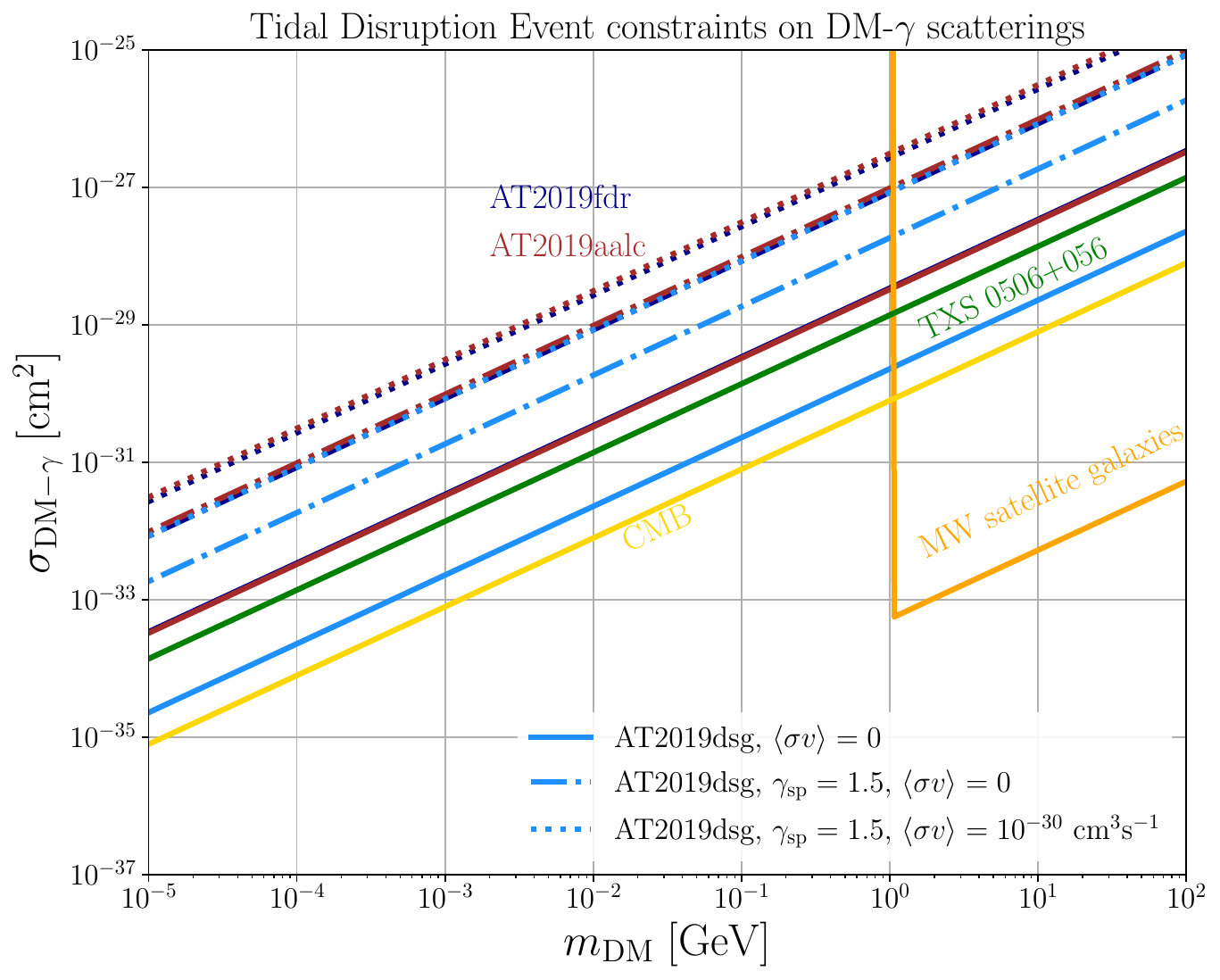}
		\centering
		\caption{\textit{Left panel}: Upper limits on the constant DM-neutrino scattering cross section from AT2019dsg, AT2019fdr and AT2019aalc (light blue, dark blue, and brown, respectively), for different assumptions on the DM distribution and astrophysical emission of neutrinos, see Table I for details. For comparison, we show previous constraints from TXS 0506+056 \cite{Ferrer:2022kei}, and Lyman-$\alpha$ \cite{Wilkinson:2014ksa}. \textit{Right panel:} Upper limits on the constant DM-photon scattering cross section. For comparison, we show previous constraints from TXS 0506+056 \cite{Ferrer:2022kei}, CMB \cite{Wilkinson:2013kia}, and Milky Way satellite galaxies \cite{Boehm:2014vja, Escudero:2015yka}}
		\label{fig:bound_TDE_constant}
	\end{figure*}

For the pre-existing profile, we will take an NFW profile \cite{Navarro:1996gj,Navarro:1995iw}, with $\gamma=1$, which evolves to a spike with $\gamma_{\rm sp}=7/3$ and $\alpha_{\gamma}\simeq 0.1$. 
In Table~\ref{tab:TDEs}, we show the masses of the central SMBHs of the three TDEs considered in this work. For the scale radius of those TDEs host galaxies, we take $r_0=10$ kpc. Finally, the normalization $\rho_0$ is determined by the uncertainty on the black hole mass \cite{Gorchtein_2010,Lacroix_2017}, also given in Table~\ref{tab:TDEs}.

In the presence of DM pair annihilation into SM particles, the maximal DM density in the spike is saturated to $\rho_{\text {sat}}= m_{\rm DM} /(\langle\sigma v \rangle t_{\mathrm{BH}})$, where $\langle \sigma v \rangle$ is the velocity averaged DM annihilation cross section, and $t_{\rm BH}$ is the black hole age. For all TDEs, we assume $t_{\rm BH}=10^{10}$ yr \cite{Piana_2020}. Moreover, the DM spike extends to a maximal radius, $R_{\rm sp}$, beyond which the DM distribution follows the previous seed profile. The DM density then reads~\cite{Gondolo_1999, Lacroix_2015, Lacroix_2017}
\begin{align}\rho(r)= \begin{cases} 
		0 & r\leq 4R_{\rm S}, \\[5pt]
		\frac{\rho_{\rm sp}(r)\rho_{\rm sat}}{\rho_{\rm sp}(r)+\rho_{\rm sat}} & 4R_{\rm S}\leq r\leq R_{\rm sp}, \\[3pt]
		\rho_{0}\left(\frac{r}{r_0}\right)^{-\gamma} \left(1+\frac{r}{r_0}\right)^{-(3-\gamma)} & r\geq R_{\rm sp} .
	\end{cases}
	\label{eq:spike_profile}
\end{align}

Besides annihilations, gravitational interactions between DM and stars surrounding the SMBH may deplete the structure of the spike. It has been shown that the spike would relax to a profile with an index as low as $\gamma_{\rm sp}=1.5$, depending on the age of the galactic bulge \cite{Bertone:2004pz}. To be conservative, we will consider this possibility in our analysis as well. In Fig.~\ref{fig:DMspike}, we show the distribution of DM particles under these assumptions on AT2019dsg, AT2019fdr, and AT2019aalc (light blue, dark blue, and brown, respectively). For comparison, we also show the region where neutrinos are likely to be emitted, according to \cite{Winter:2022fpf}.
The solid, dashed, and dotted curves represent uncertainty in DM density profiles, which are defined as (1), (2), and (3) in Table~\ref{tab:TDEs}, respectively. Profile (1) is the optimized profile with a large DM density spike ($\gamma_{\rm sp}  =  2.25$, and assuming no DM annihilation), and we also take the closest emission radius within the uncertainty range. As we will see below, this scenario gives the most stringent bound. In profile (2), we still assume no DM annihilation and the closest emission radius but take a spike profile depleted by interactions with stars, $\gamma_{\rm sp}  =  1.5$. Profile (3) is the most conservative choice of parameters, where we consider DM self-annihilations with $\langle \sigma  v\rangle/m_{\rm 
 DM}  =  10^{-30}~\mathrm{cm}^3  \mathrm{s}^{-1}  / \mathrm{GeV}$.

\section{Upper limits on DM-neutrino and DM-photon scatterings}\label{sec:UpperLimits}

\begin{figure*}[t!]
    \centering
\includegraphics[width=0.49\textwidth]{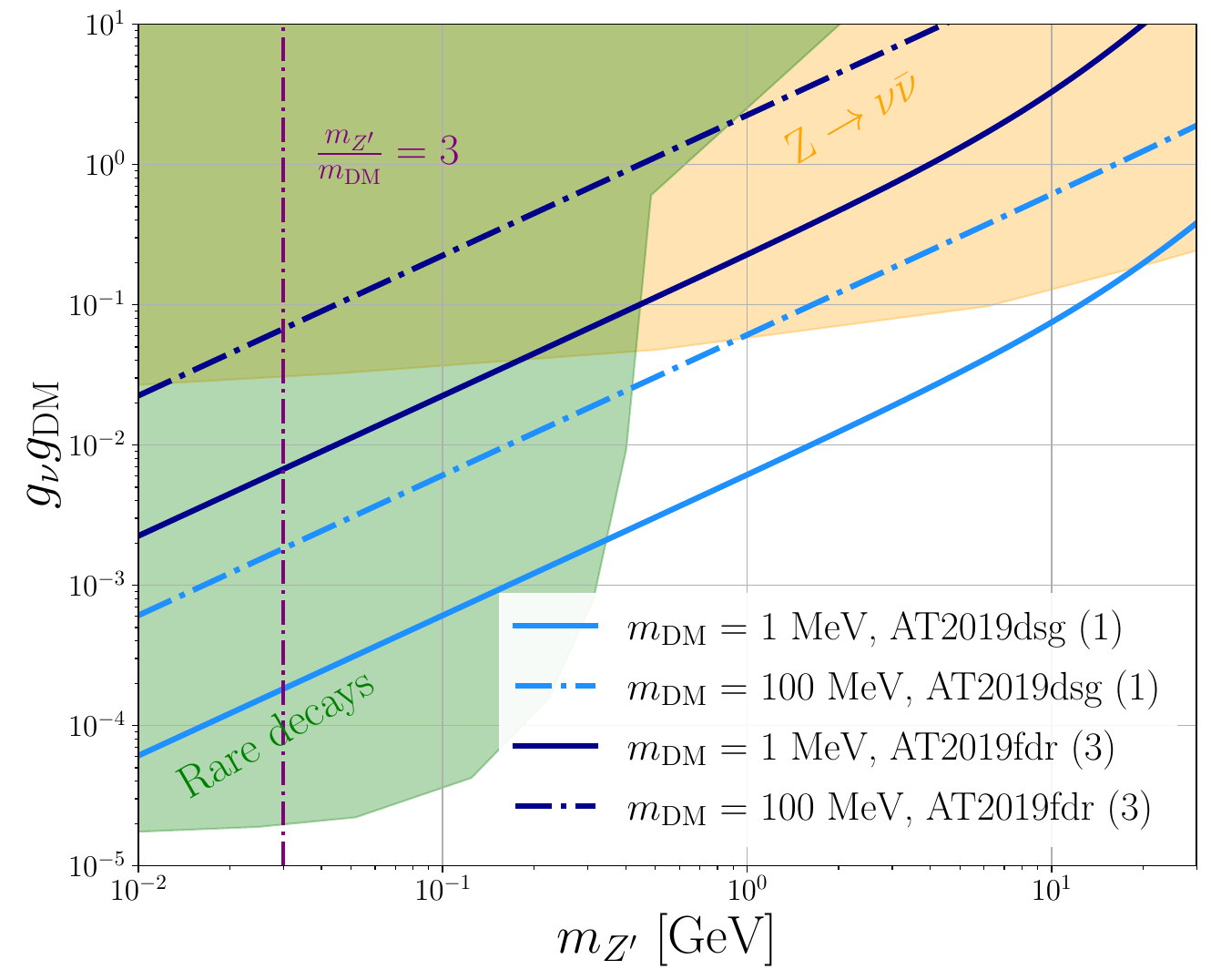}
\includegraphics[width=0.49\textwidth]{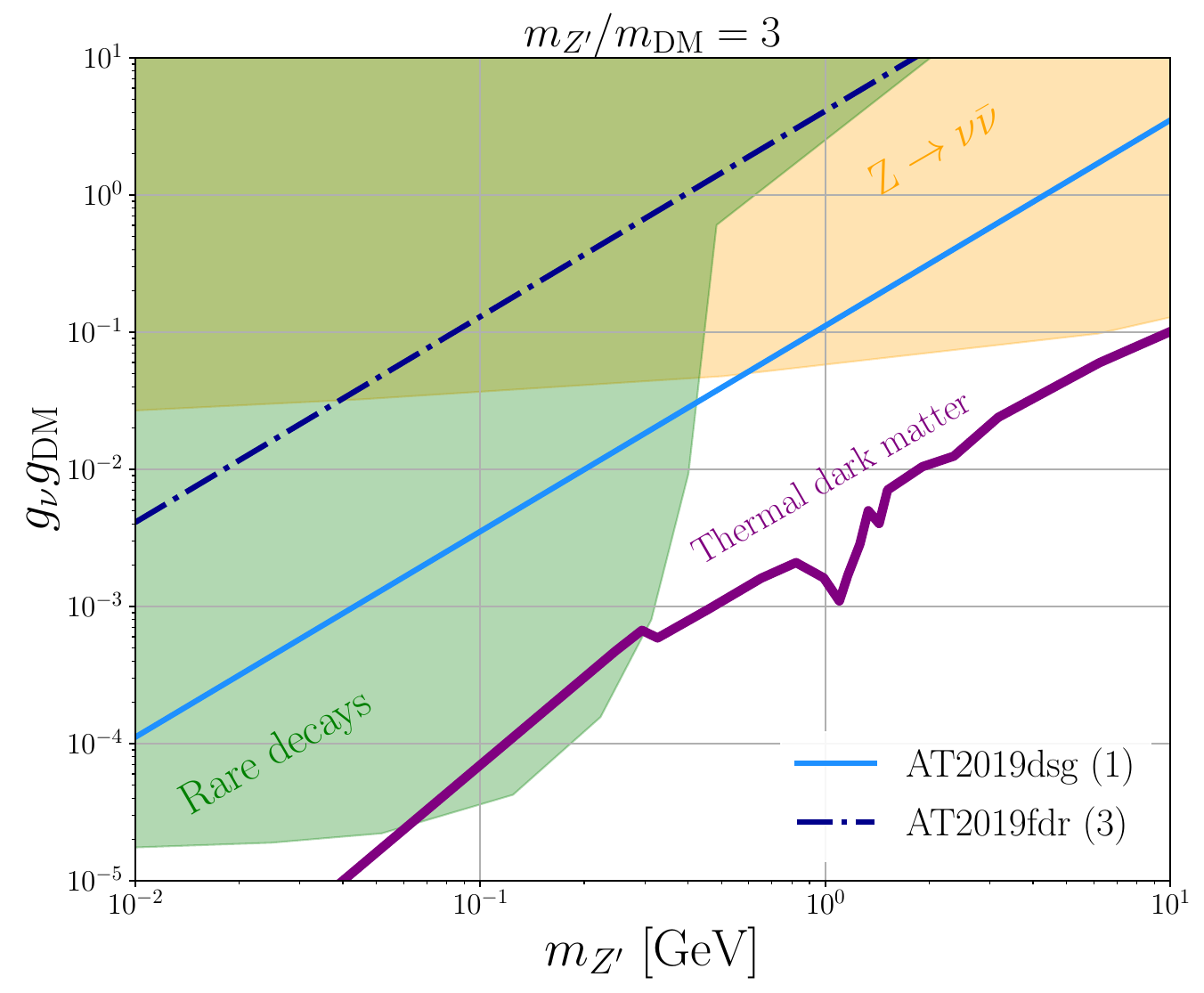}
    \caption{\textit{Left}: Constraints in the product of gauge couplings $g_{\nu}g_{\rm DM}$ vs mediator mass $m_{Z^{\prime}}$, for fixed values of the DM mass, and from the most stringent TDE (AT2019dsg (1), light blue) and least stringent (AT2019dfr (3), darkblue) considered in this work. For comparison, we show complementary constraints from invisible decays of $Z$ boson and rare decay processes of pions and kaons \cite{Berryman:2022hds}. We further show as a vertical dashed-dotted purple line a constraint on thermal dark matter for $m_{Z^{\prime}}/m_{\rm DM}=3$ and $m_{\rm DM}=100$ MeV, for comparison with the plot on the right. The region shown in the plot is fully allowed for $m_{\rm DM}=1$ MeV, and thus not displayed. 
    \textit{Right:} Constraints in the product of gauge couplings $g_{\nu}g_{\rm DM}$ vs mediator mass ${Z^{\prime}}$, for $m_{Z^{\prime}}/m_{\rm DM}=3$ most stringent TDE (AT2019dsg (1), light blue) and least stringent (AT2019dfr (3), darkblue) considered in this work. For comparison, we show values of couplings corresponding to thermal DM in a $Z^{\prime}$ model \cite{Berlin:2018bsc}.}
\label{fig:bound_TDE_Zprime}
\end{figure*}

\subsection{Constraint on constant cross section}

In this work, we consider different interaction models between DM and neutrinos/photons. First, we will derive upper limits assuming that the DM-neutrino and the DM-photon scattering cross section is constant. The flux of a particle $i=\nu, \gamma$ evolves due to interactions with DM particles via the cascade equation
\begin{align}
\frac{d \Phi}{d \tau}
&=  -\sigma_{{\rm DM}\text{-}i} {\Phi}
\nonumber
\\
&~~~~
 +\int_{E_{i}}^{\infty} d E_{i}^{\prime} \frac{d \sigma_{{\rm DM}\text{-}i}}{d E_{i}}\left(E_{i}^{\prime} \rightarrow E_{i}\right)\Phi\left(E_{i}^{\prime}\right),
 \label{eq:cascade}
\end{align}
where $\tau=\Sigma_{\rm DM}/m_{\rm DM}$, and $\Sigma_{\rm DM}$ is the column density of DM particles along the line of sight (l.o.s) between the TDE and the Earth, which is largely dominated by the contributions from the spike and the halo of the host galaxy
\begin{align}
  \Sigma_{\rm  DM}  \equiv  \int_{\rm  l.o.s}  dr \rho  (r) = \left.\Sigma_{\mathrm{DM}}\right|_{\rm spike}+\left.\Sigma_{\mathrm{DM}}\right|_{\rm halo}.
\end{align}
since the column density in the intergalactic medium and the Milky Way is much smaller. The farthest TDE considered in this work is AT2019fdr, with redshift $z=0.267$ \cite{Winter:2022fpf}. We find that the column density of DM in the intergalactic medium, from AT2019fdr, barely exceeds $ \sim 10^{21}$ GeV/cm$^2$. As for the Milky Way, provided the high-energy neutrinos from the TDEs do not come from the direction of the galactic center, we find that the column density hardly exceeds $\sim 10^{24}$ GeV/cm$^2$.

If the cross section is constant, the second term of the cascade equation can be neglected, and we obtain a simple exponential factor controlling the attenuation of the fluxes. 
The IceCube collaboration has yet not claimed an association of high-energy neutrinos with the TDEs considered here, thus, we cannot assess precisely what reduction in the expected emitted neutrino fluxes (\textit{E.g} from \cite{Winter:2022fpf}) would contradict observations. In this paper, to reveal the impact of potential attenuation at TDEs qualitatively, we impose that the attenuation of the neutrino flux shall not be larger than $90\%$ of the initial emitted neutrino flux, and that the attenuation of the photon flux should not be larger than $99\%$ of the initial emitted flux. 
Notice that photon flux experiences more attenuation from the SM particles compared to neutrino flux. This is why we require more severe criteria (99\% instead of 90\%) to derive constraints on DM-photon interaction.
In this case, the upper limits are given by \cite{Ferrer:2022kei}:

\begin{align}
		\frac{\sigma_{\rm DM\text{-}\nu}}{m_{\mathrm{DM}}} & \lesssim \frac{2.3}{\Sigma_{\rm DM}},\;
		\label{eq:criteriaConstantUL}
	\end{align}
 and
\begin{align}
		\frac{\sigma_{\rm DM\text{-}\gamma}}{m_{\mathrm{DM}}} & \lesssim \frac{4.6}{\Sigma_{\rm DM}}.\;
	\end{align}
We notice that $\Sigma_{\rm  DM}$ is largely dominated by the contribution from the spike. The halo of the host galaxy may contribute significantly if neutrinos and photons are emitted at $R_{\rm em} \gtrsim 10^4 R_{\rm  S}$. These can be calculated analytically in the absence of annihilations as \cite{Ferrer:2022kei}
\begin{equation}
\left.\Sigma_{\mathrm{DM}}\right|_{\mathrm{spike}} \simeq \frac{\rho_{\mathrm{sp}}\left(R_{\mathrm{em}}\right) R_{\mathrm{em}}}{\left(\gamma_{\mathrm{sp}}-1\right)}\left[1-\left(\frac{R_{\mathrm{sp}}}{R_{\mathrm{em}}}\right)^{1-\gamma_{\mathrm{sp}}}\right],
\label{eq:column_density-approximated-spike}
\end{equation}
and
\begin{equation}
\left.\Sigma_{\mathrm{DM}}\right|_{\text {halo }}\simeq \rho_0 r_0\left[\log \left(\frac{r_0}{R_{\mathrm{sp}}}\right)-1\right].
\label{eq:column_density-approximated-halo}
\end{equation}
where the former (latter) formula assumes $R_{\rm  em} \ll  R_{\rm sp}$ ($r_0 \gg R_{\mathrm{sp}}$). These approximations are valid for our models (1) and (2), while for model (3), we obtain the column density numerically. We note that the spike contribution is proportional to $\rho_{\rm sp}  (R_{\rm  em})$, which mainly characterizes the DM column density. 

In Fig. \ref{fig:bound_TDE_constant}, we show constraints on the DM-neutrino and DM-photon scattering cross sections, for the different TDEs considered in this work, and the different set of assumptions discussed in Table~\ref{tab:TDEs}. The strength of our constraints can vary for different astrophysical models and TDE sources within $\sim 4$ orders of magnitude. Our most optimistic scenario yields a constraint on the DM-neutrino scattering cross section which is $\sim 1$ order of magnitude weaker than the constraint from Lyman-$\alpha$ \cite{Wilkinson:2014ksa}, and it is a factor of $\sim$ 4 stronger than the constraint obtained from TXS 0506+056 \cite{Ferrer:2022kei}.

The spike contribution on DM column density is characterized by the value of $\rho_{\rm sp}  (R_{\rm  em})$ (see Eq.~\eqref{eq:column_density-approximated-spike}). AT2019dsg realizes the most stringent constraint among the three TDEs, and thus we obtain the highest value of $\rho_{\rm sp}  (R_{\rm  em})$ as listed in the last column in Table~\ref{tab:TDEs}.

For DM-photon scatterings, our most optimistic scenario gives a constraint that is weaker than the bound from the Cosmic Microwave Background \cite{Wilkinson:2013kia} by a factor of $2$ for DM masses below $m_{\rm DM} \lesssim 1$ GeV, and weaker than constraints from Milky Way satellite galaxies at larger masses~\cite{Escudero:2015yka} by $\sim 3$ orders of magnitude.

\subsection{Constraint on simplified models}

Now we consider as a simplified model a Dirac fermion composing the DM of the Universe, that interacts with neutrinos and photons via a vector mediator. We find that the scattering cross section with neutrinos, and the differential scattering cross section for a neutrino to go from energy $E_{\nu}$ to an energy $E_{\nu}^{\prime}$, are given by
\begin{widetext}
\begin{align}
    \sigma_{\rm DM\text{-}\nu} 
    &
    = \frac{g_{\nu}^2 g_{\rm DM}^2}{16 \pi E_\nu^2 m_{\rm DM}^2}
    \Bigl[ 
    (2 E_\nu m_{\rm DM} + m_{{Z^{\prime}}}^2 + m_{\rm DM}^2) \log \left( \frac{m_{Z^{\prime}}^2 (2 E_\nu + m_{\rm DM})}{4 E_\nu^2 m_{\rm DM} + 2 E_\nu m_{{Z^{\prime}}}^2 + m_{\rm DM} m_{{Z^{\prime}}}^2} \right) 
    \nonumber
    \\
    &~~~~~~~~~~~~~~~~~~~~~~
    + 4 E_\nu^2 \left( 1 + \frac{m_{\rm DM}^2}{m_{{Z^{\prime}}}^2} - \frac{2 E_\nu (4 E_\nu^2 m_{\rm DM} + E_\nu (m_{\rm DM}^2 + 2 m_{{Z^{\prime}}}^2) + m_{\rm DM} m_{{Z^{\prime}}}^2)}{(2 E_\nu + m_{\rm DM}) (m_{\rm DM} (4 E_\nu^2 + m_{{Z^{\prime}}}^2) + 2 E_\nu m_{{Z^{\prime}}}^2)} \right) \Bigr],
    \\
    \frac{d \sigma_{{\rm DM}-\nu}^{E_{\nu} \rightarrow E_{\nu}^{\prime}}}{d E_\nu} 
    &
    = \frac{g_\nu^2 g_{\rm DM}^2 m_{\rm DM}(5 E_\nu^2 + E_\nu (m_{\rm DM} - 4 E_\nu^{\prime}) + E_\nu^{\prime}(E_\nu^{\prime} - m_{\rm DM}))}{16 \pi E_\nu^2 (2 m_{\rm DM}(E_\nu^{\prime} - E_\nu) + m_{{Z^{\prime}}}^2)^2}.
\end{align}
 \end{widetext}

With these ingredients, we can solve the cascade equation for the predicted neutrino fluxes from \cite{Winter:2022fpf}. We use the model corresponding to high-energy neutrinos being produced by moderate-energy protons interacting with ambient X-rays, see Fig. \ref{fig:fluences} for the time-integrated neutrino fluxes (fluences) from the TDEs considered in this work \cite{Winter:2022fpf}. We have checked that our results do not change significantly when using the expected fluxes predicted in the model with medium-energy protons interacting with OUV photons, or when using the model with high-energy protons interacting with IR photons.

Now we can derive constraints on the parameter space of the DM mass, mediator mass and couplings from the consideration that the neutrino fluxes shall not be suppressed more than 90$\%$ at the claimed detected neutrino energies from AT2019dsg (217 TeV), AT2019fdr (83 TeV) and AT2019aalc (176 TeV). We show our results in Fig. \ref{fig:bound_TDE_Zprime}, for DM masses of 1--100 MeV, and from the most aggressive and conservative models of the TDEs considered in this work, which are AT2019dsg (1) and AT2019fdr(3), respectively. The column density along the line of sight of AT2019fdr(3) is smaller than that of AT2019aalc(3) by 10$\%$, and thus AT2019fdr(3) gives a slightly weaker constraint than that of AT2019aalc(3) in Fig.~\ref{fig:bound_TDE_constant}.
However, due to the smaller energy of the neutrinos detected, energy-dependent cross section constraints on AT2019fdr(3) will be weaker than for AT2019aalc(3). We further show complementary constraints in the shaded yellow region from invisible decays of $Z$ boson \cite{Berryman:2018ogk, deGouvea:2019qaz}, and in the green region from rare decay processes of pions and kaons \cite{Laha:2013xua,Ibe:2016dir,Bakhti:2017jhm}. At lower DM masses than shown in the plot, constraints from double beta decay and BBN also apply \cite{Burgess:1992dt,Brune:2018sab,Cepedello:2018zvr,Huang:2017egl,Grohs:2020xxd}. We note that some of these complementary constraints only apply to scalar mediators. For vector mediators, there are additional constraints from neutrino-electron scattering in the limit where the mediator couples only to neutrinos \cite{Chauhan:2022iuh}.
Our constraints from AT2019dsg (1) are stronger than previous ones for mediators lighter than a few GeV, while they are only stronger in some regions of parameter space in the model AT2019fdr (3), and only when the DM mass is lighter than $\sim 1$~MeV.

Let us comment on the complementarity of our constraints with those values of the DM-neutrino scattering cross section able to explain some cosmological tensions and small-scale structure anomalies, \textit{E.g} \cite{Bertoni:2014mva,DiValentino:2017oaw,Olivares-DelCampo:2017feq,Hooper:2021rjc,Brax:2023tvn,Brax:2023rrf,Akita:2023yga,Giare:2023qqn}. These studies yield values of the DM-neutrino scattering cross section of $\sigma_{\rm DM\text{-}\nu} \simeq 10^{-31}$--$10^{-30}$~cm$^2$. Such small values can only be probed in the constant cross section scenario in our AT2019dsg (1) model.

\begin{figure}[tb]
    \centering
\includegraphics[width=0.49\textwidth]{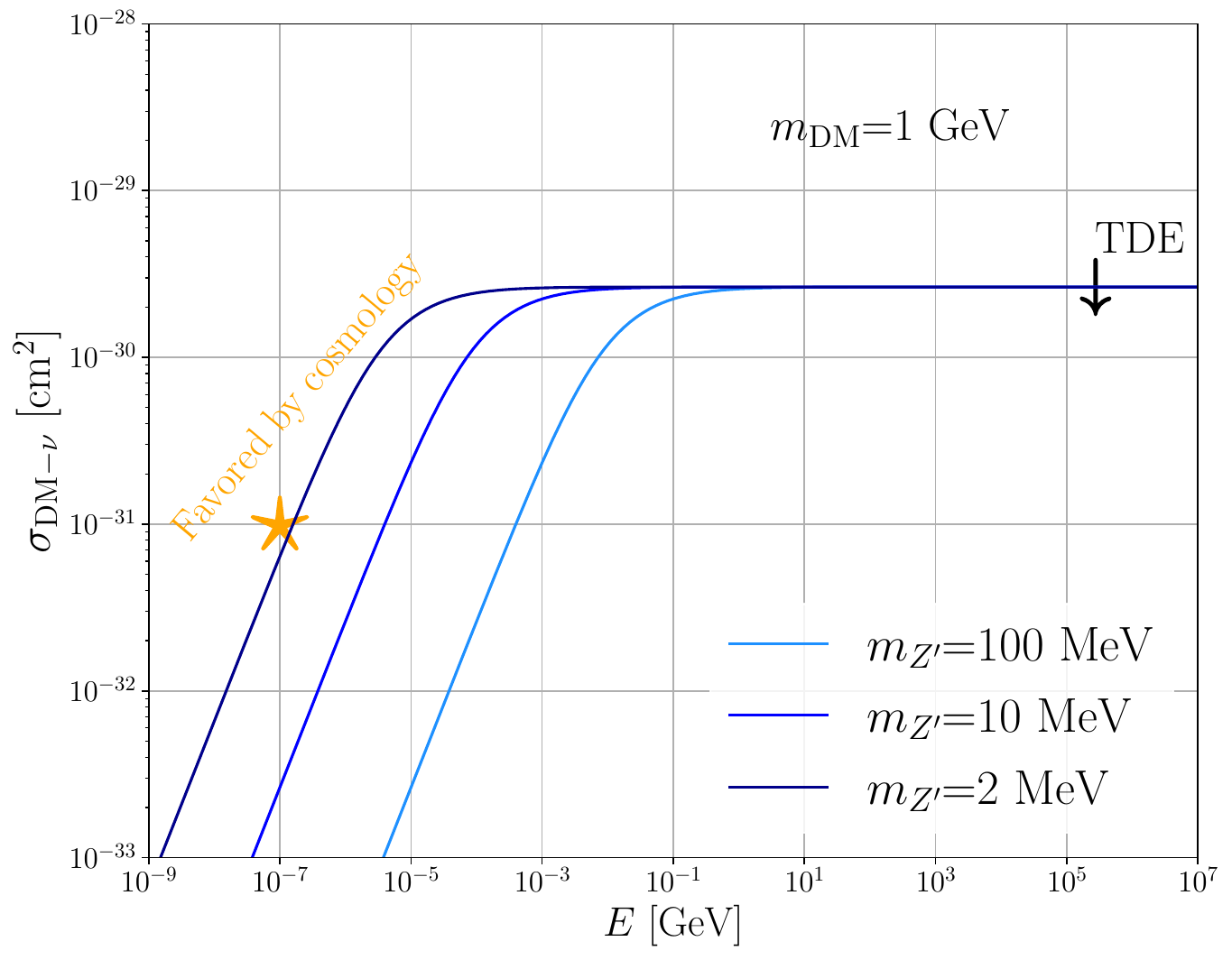}
    \caption{Comparison of our constraints on the DM neutrino scattering cross section from the model AT2019dsg (3) with the preferred values of the DM-neutrino scattering cross section able to explain cosmological tensions. For a DM mass of $m_{\rm DM}=1$ GeV and mediators heavier than $m_{Z^{\prime}} \gtrsim 2$ MeV, our extrapolated constraints rule out the cosmological values.}
\label{fig:cosmo_vs_TDE}
\end{figure}

On the other hand, in models where the DM neutrino cross-section rises with energy, the favored cosmological values of the DM-neutrino cross section will be in conflict with most of our TDE models, since the neutrino energies from the TDE are of order $\sim 100$ TeV, while the cosmological observables hardly probe energies larger than $\sim 100$ eV. Fig. \ref{fig:cosmo_vs_TDE} indicates that the complementarity of cosmological observables and astrophysics can be relevant even for light mediators; for a DM mass of $m_{\rm DM}=$1 GeV, and mediators heavier than $m_{Z^{\prime}} \gtrsim 2$ MeV, TDEs may probe the values favored by cosmology.

Now we turn into signatures of DM-photon scattering within the same model that we used for DM-neutrino interactions. Fermionic DM can also scatter off photons via a vector mediator in a Compton-like inelastic scattering process, due to the kinetic mixing between the new gauge boson and the SM photon. The inelastic process $\mathrm{DM} \gamma \rightarrow \mathrm{DM} {Z^{\prime}}$ dominates, with ${Z^{\prime}}$ the vector boson in the final state. Here, we focus on the limit where the DM particle is much heavier than the relevant photon energies (X-rays) and the vector mediator mass $m_{\rm DM} \gg E_\gamma \gg m_{Z^{\prime}}$. We find that the cross section can be expressed to a good approximation as
\begin{equation}
\sigma_{\mathrm{DM}-\gamma} \simeq \frac{\alpha\epsilon^2 g_{\rm DM}^2 \sqrt{E_{\gamma}^2-m_{Z^{\prime}}^2}\left(2E_{\gamma}^2+m_{Z^{\prime}}^2\right)}{3E_{\gamma}^3 m_{\rm DM}^2},
\end{equation}
where $\alpha$ is the electromagnetic fine structure constant, $g_{\rm DM}$ is the dark boson coupling to the DM fermion, and $\epsilon$ is the kinetic mixing between the vector boson and the SM photon. Since the leading order contribution is the inelastic scattering channel, the second term in the cascade equation vanishes, and the attenuation of the fluxes is exponential.

\begin{figure}[htb]
    \centering
\includegraphics[width=0.49\textwidth]{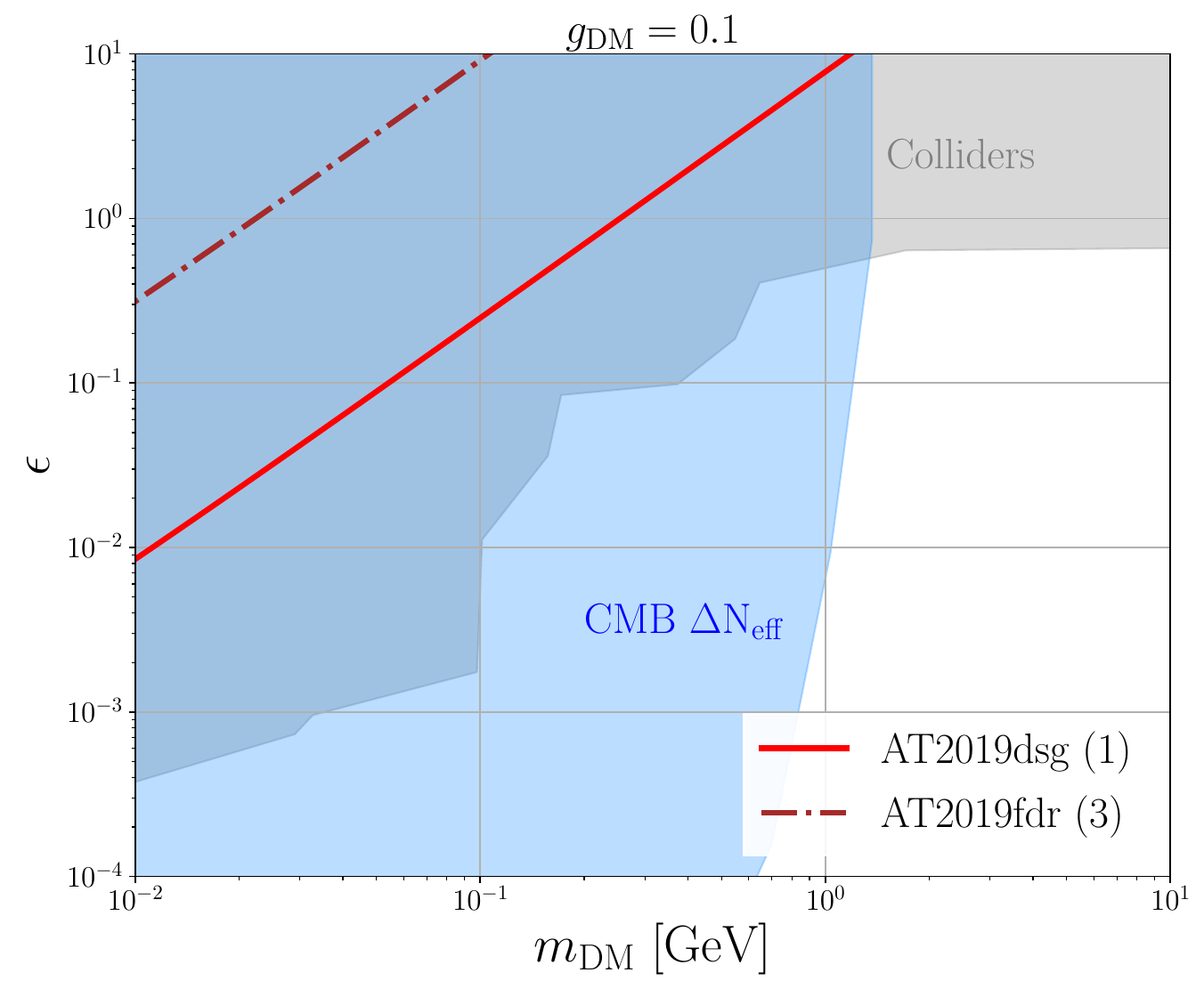}
    \caption{Constraints in the kinetic mixing parameter $\epsilon$ vs DM mass, for light mediators $ m_{\rm DM} \gg m_{Z^{\prime}}$, from the most stringent TDE (AT2019dsg (1), 
    red solid
    ) and least stringent (AT2019dfr (3), 
    brown dotdashed
    ) considered in this work. For comparison, we show complementary constraints from cosmological determinations of $N_{\rm eff}$ at the time of BBN and CMB \cite{Vogel:2013raa} (relevant at $m_{\rm DM} \lesssim$ 1 GeV) and from colliders (relevant at $m_{\rm DM} \lesssim$ 1 GeV) \cite{Davidson:2000hf, Liu:2019knx}.}
\label{fig:bound_TDE_Zprime_photon}
\end{figure}

In Fig. \ref{fig:bound_TDE_Zprime_photon}, we show constraints on the kinetic mixing between the vector mediator and the SM photon, as a function of the DM mass, and in the limit where the mediator is ultralight or massless $m_{\rm DM} \gg m_{\rm Z^{\prime}}$. For this purpose, we impose the 99$\%$ attenuation criteria on the emitted fluxes from AT2019dsg(1) and AT2019fdr(3). We find our constraint by varying the photon energy in the measured X-ray ranges, which goes from 0.3~keV to 10~keV for AT2019dsg \cite{Stein:2020xhk}, and from 0.3~keV to 2~keV for AT2019fdr \cite{Reusch:2021ztx}.

For this concrete model of DM-photon interactions, we are not aware of any other astrophysical constraint for $m_{\rm DM} \gtrsim 10$~MeV. However, they are weaker than the combination of cosmological and collider constraints when fixing the dark gauge coupling to the canonical value $g_{\rm DM}=0.1$. Astrophysical constraints could become competitive with cosmology and colliders if the gauge coupling is saturated to the perturbative limit $g_{\rm DM}\sim 1$--$10$, or if the column density of DM particles is at least one order of magnitude larger than in the TDEs considered in this work, $\Sigma_{\rm DM} \gtrsim 2 \times 10^{31}$ GeV/cm$^2$.

\section{Discussion and conclusions}
\label{sec:conclusions}
Multimessenger emission of TDEs has been poorly studied in the context of BSM physics. Here, we explored the potential attenuation of the emitted neutrino and photon fluxes from TDEs due to scatterings with DM particles around the source. A clear advantage of TDEs to probe these models is their emission region, which is determined to be very close to the SMBH, where there might be a high DM density region called the DM spike. Various directions have been proposed to exploit the attenuation of standard model particles within DM spikes, focusing on AGNs. We, for the first time, propose to use TDEs to probe the DM scattering properties.

The DM column density along the line of sight from the source to the observed point is crucial to have observable attenuation effects, which are dominated by the DM density profile around the host SMBH. This column density is proportional to the DM density at the emission region of the source, $\rho_{\rm  DM} (R_{\rm em})$ (see Eq.~\eqref{eq:column_density-approximated-spike}). As listed in Table~\ref{tab:TDEs}, the highest density is obtained for AT2019dsg, due to its smaller black hole mass (see Fig.~\ref{fig:DMspike}). The derived constraint is, therefore, the most stringent compared to other TDE candidates. For constant DM-neutrino (DM-photon) scattering cross sections, we find that AT2019dsg gives a stronger constraint than the blazar TXS 0560+056 by a factor of $\sim  4 (2)$.

We further discussed more realistic constraints using a simplified model of DM, where a $Z'$ boson mediates the DM interaction with the neutrino and photon. For the neutrino channel, the cross section is increased in the high energy limit. The TDE candidates considered in our work are detected in neutrinos of $\sim  100~\mathrm{TeV}$ energies, and have great advantages in probing scattering effects in models with heavy mediators. Furthermore, we discuss how TDE observations may probe the values of the DM-neutrino scattering cross section favored to address some cosmological tensions $\sigma_{{\rm  DM\text{-}}\nu}  \sim  10^{-31}$--$10^{-30}~\mathrm{cm}^2$. Using AT2019dsg, we found that such values can be probed in models where $m_{\rm  DM}  \gtrsim  2~\mathrm{MeV}$ (see Fig.~\ref{fig:cosmo_vs_TDE}). For DM-photon scatterings via a light vector boson, the leading order process is the inelastic scattering DM $\gamma  \to$ DM $Z'$ due to the kinetic mixing suppression of the elastic channel, and thus the attenuation factor again has a simple exponential form. We found that current TDE candidates give weaker constraints than other complementary constraints from cosmology and colliders. Future observations of TDEs may improve the current constraint, which lies roughly an order of magnitude weaker than collider bounds around $m_{\rm DM} \sim 1$ GeV. If we observe TDEs from black holes with with smaller masses than considered here, we may obtain stronger constraints, due to an enhancement in the DM column density. Besides, it is essential to narrow down the uncertainties from the emission region and mechanisms of TDEs to improve our constraints.

In this work, we did not consider the time information of TDEs but, more intriguingly, the hints of high-energy neutrino emission from AT2019dsg, AT2019fdr, and AT2019aalc occurred $\mathcal{O}(100)$ days after the maximum of the optical-ultraviolet luminosity. The possibility has been raised that the delay in neutrino arrival time may be connected to how the mass accretion rate around the SMBH changes over time. This change could be influenced by the circularization time of debris or could remain consistent for several hundred days \cite{vanVelzen:2021zsm}. This interpretation is yet to be confirmed with further sources, and requires more complex modeling of the underlying physics at the source, since hadronic processes produce both neutrinos and photons simultaneously via pion decay. While the delay in neutrino arrival may be related to the characteristic timescale of the system, or just a statistical effect, it may also be induced by DM-neutrino interactions at the source. We will address this possibility in detail in our upcoming paper.

\section{Acknowledgments}
We are grateful to Francesca Capel, Chengchao Yuan, Walter Winter and Kohta Murase for discussions on Tidal Disruption Events, to Alejandro Ibarra for discussions on models of dark matter-photon interactions, and to Mathias Garny for conversations on dark matter-neutrino interactions in cosmology. GH thanks DESY Zeuthen for hospitality during his visit. This work was supported by the Collaborative Research Center SFB1258 and by the Deutsche Forschungsgemeinschaft (DFG, German Research Foundation) under Germany's Excellence Strategy - EXC-2094 - 390783311. The work of GH was also supported by the U.S. Department of Energy Office of Science under award number DE-SC0020262.

\appendix

\bibliographystyle{apsrev4-1}
\bibliography{References}

\begin{thebibliography}{114}%
\makeatletter
\providecommand \@ifxundefined [1]{%
 \@ifx{#1\undefined}
}%
\providecommand \@ifnum [1]{%
 \ifnum #1\expandafter \@firstoftwo
 \else \expandafter \@secondoftwo
 \fi
}%
\providecommand \@ifx [1]{%
 \ifx #1\expandafter \@firstoftwo
 \else \expandafter \@secondoftwo
 \fi
}%
\providecommand \natexlab [1]{#1}%
\providecommand \enquote  [1]{``#1''}%
\providecommand \bibnamefont  [1]{#1}%
\providecommand \bibfnamefont [1]{#1}%
\providecommand \citenamefont [1]{#1}%
\providecommand \href@noop [0]{\@secondoftwo}%
\providecommand \href [0]{\begingroup \@sanitize@url \@href}%
\providecommand \@href[1]{\@@startlink{#1}\@@href}%
\providecommand \@@href[1]{\endgroup#1\@@endlink}%
\providecommand \@sanitize@url [0]{\catcode `\\12\catcode `\$12\catcode `\&12\catcode `\#12\catcode `\^12\catcode `\_12\catcode `\%12\relax}%
\providecommand \@@startlink[1]{}%
\providecommand \@@endlink[0]{}%
\providecommand \url  [0]{\begingroup\@sanitize@url \@url }%
\providecommand \@url [1]{\endgroup\@href {#1}{\urlprefix }}%
\providecommand \urlprefix  [0]{URL }%
\providecommand \Eprint [0]{\href }%
\providecommand \doibase [0]{http://dx.doi.org/}%
\providecommand \selectlanguage [0]{\@gobble}%
\providecommand \bibinfo  [0]{\@secondoftwo}%
\providecommand \bibfield  [0]{\@secondoftwo}%
\providecommand \translation [1]{[#1]}%
\providecommand \BibitemOpen [0]{}%
\providecommand \bibitemStop [0]{}%
\providecommand \bibitemNoStop [0]{.\EOS\space}%
\providecommand \EOS [0]{\spacefactor3000\relax}%
\providecommand \BibitemShut  [1]{\csname bibitem#1\endcsname}%
\let\auto@bib@innerbib\@empty
\bibitem [{\citenamefont {{Hills}}(1975)}]{1975Natur.254..295H}%
  \BibitemOpen
  \bibfield  {author} {\bibinfo {author} {\bibfnamefont {J.~G.}\ \bibnamefont {{Hills}}},\ }\href {\doibase 10.1038/254295a0} {\bibfield  {journal} {\bibinfo  {journal} {\nat}\ }\textbf {\bibinfo {volume} {254}},\ \bibinfo {pages} {295} (\bibinfo {year} {1975})}\BibitemShut {NoStop}%
\bibitem [{\citenamefont {Rees}(1988)}]{Rees:1988bf}%
  \BibitemOpen
  \bibfield  {author} {\bibinfo {author} {\bibfnamefont {M.~J.}\ \bibnamefont {Rees}},\ }\href {\doibase 10.1038/333523a0} {\bibfield  {journal} {\bibinfo  {journal} {Nature}\ }\textbf {\bibinfo {volume} {333}},\ \bibinfo {pages} {523} (\bibinfo {year} {1988})}\BibitemShut {NoStop}%
\bibitem [{\citenamefont {Komossa}(2015)}]{Komossa:2015qya}%
  \BibitemOpen
  \bibfield  {author} {\bibinfo {author} {\bibfnamefont {S.}~\bibnamefont {Komossa}},\ }\href {\doibase 10.1016/j.jheap.2015.04.006} {\bibfield  {journal} {\bibinfo  {journal} {JHEAp}\ }\textbf {\bibinfo {volume} {7}},\ \bibinfo {pages} {148} (\bibinfo {year} {2015})},\ \Eprint {http://arxiv.org/abs/1505.01093} {arXiv:1505.01093 [astro-ph.HE]} \BibitemShut {NoStop}%
\bibitem [{\citenamefont {{Murase}}(2008)}]{2008AIPC.1065..201M}%
  \BibitemOpen
  \bibfield  {author} {\bibinfo {author} {\bibfnamefont {K.}~\bibnamefont {{Murase}}},\ }in\ \href {\doibase 10.1063/1.3027912} {\emph {\bibinfo {booktitle} {2008 Nanjing Gamma-ray Burst Conference}}},\ \bibinfo {series} {American Institute of Physics Conference Series}, Vol.\ \bibinfo {volume} {1065},\ \bibinfo {editor} {edited by\ \bibinfo {editor} {\bibfnamefont {Y.-F.}\ \bibnamefont {{Huang}}}, \bibinfo {editor} {\bibfnamefont {Z.-G.}\ \bibnamefont {{Dai}}}, \ and\ \bibinfo {editor} {\bibfnamefont {B.}~\bibnamefont {{Zhang}}}}\ (\bibinfo {year} {2008})\ pp.\ \bibinfo {pages} {201--206}\BibitemShut {NoStop}%
\bibitem [{\citenamefont {Wang}\ \emph {et~al.}(2011{\natexlab{a}})\citenamefont {Wang}, \citenamefont {Liu}, \citenamefont {Dai},\ and\ \citenamefont {Cheng}}]{Wang_2011}%
  \BibitemOpen
  \bibfield  {author} {\bibinfo {author} {\bibfnamefont {X.-Y.}\ \bibnamefont {Wang}}, \bibinfo {author} {\bibfnamefont {R.-Y.}\ \bibnamefont {Liu}}, \bibinfo {author} {\bibfnamefont {Z.-G.}\ \bibnamefont {Dai}}, \ and\ \bibinfo {author} {\bibfnamefont {K.~S.}\ \bibnamefont {Cheng}},\ }\href {\doibase 10.1103/physrevd.84.081301} {\bibfield  {journal} {\bibinfo  {journal} {Physical Review D}\ }\textbf {\bibinfo {volume} {84}} (\bibinfo {year} {2011}{\natexlab{a}}),\ 10.1103/physrevd.84.081301}\BibitemShut {NoStop}%
\bibitem [{\citenamefont {Farrar}\ and\ \citenamefont {Piran}(2014)}]{Farrar:2014yla}%
  \BibitemOpen
  \bibfield  {author} {\bibinfo {author} {\bibfnamefont {G.~R.}\ \bibnamefont {Farrar}}\ and\ \bibinfo {author} {\bibfnamefont {T.}~\bibnamefont {Piran}},\ }\href@noop {} {\  (\bibinfo {year} {2014})},\ \Eprint {http://arxiv.org/abs/1411.0704} {arXiv:1411.0704 [astro-ph.HE]} \BibitemShut {NoStop}%
\bibitem [{\citenamefont {Pfeffer}\ \emph {et~al.}(2017)\citenamefont {Pfeffer}, \citenamefont {Kovetz},\ and\ \citenamefont {Kamionkowski}}]{Pfeffer:2015idq}%
  \BibitemOpen
  \bibfield  {author} {\bibinfo {author} {\bibfnamefont {D.~N.}\ \bibnamefont {Pfeffer}}, \bibinfo {author} {\bibfnamefont {E.~D.}\ \bibnamefont {Kovetz}}, \ and\ \bibinfo {author} {\bibfnamefont {M.}~\bibnamefont {Kamionkowski}},\ }\href {\doibase 10.1093/mnras/stw3337} {\bibfield  {journal} {\bibinfo  {journal} {Mon. Not. Roy. Astron. Soc.}\ }\textbf {\bibinfo {volume} {466}},\ \bibinfo {pages} {2922} (\bibinfo {year} {2017})},\ \Eprint {http://arxiv.org/abs/1512.04959} {arXiv:1512.04959 [astro-ph.HE]} \BibitemShut {NoStop}%
\bibitem [{\citenamefont {Dai}\ and\ \citenamefont {Fang}(2017)}]{Dai:2016gtz}%
  \BibitemOpen
  \bibfield  {author} {\bibinfo {author} {\bibfnamefont {L.}~\bibnamefont {Dai}}\ and\ \bibinfo {author} {\bibfnamefont {K.}~\bibnamefont {Fang}},\ }\href {\doibase 10.1093/mnras/stx863} {\bibfield  {journal} {\bibinfo  {journal} {Mon. Not. Roy. Astron. Soc.}\ }\textbf {\bibinfo {volume} {469}},\ \bibinfo {pages} {1354} (\bibinfo {year} {2017})},\ \Eprint {http://arxiv.org/abs/1612.00011} {arXiv:1612.00011 [astro-ph.HE]} \BibitemShut {NoStop}%
\bibitem [{\citenamefont {Lunardini}\ and\ \citenamefont {Winter}(2017)}]{Lunardini:2016xwi}%
  \BibitemOpen
  \bibfield  {author} {\bibinfo {author} {\bibfnamefont {C.}~\bibnamefont {Lunardini}}\ and\ \bibinfo {author} {\bibfnamefont {W.}~\bibnamefont {Winter}},\ }\href {\doibase 10.1103/PhysRevD.95.123001} {\bibfield  {journal} {\bibinfo  {journal} {Phys. Rev. D}\ }\textbf {\bibinfo {volume} {95}},\ \bibinfo {pages} {123001} (\bibinfo {year} {2017})},\ \Eprint {http://arxiv.org/abs/1612.03160} {arXiv:1612.03160 [astro-ph.HE]} \BibitemShut {NoStop}%
\bibitem [{\citenamefont {Senno}\ \emph {et~al.}(2017)\citenamefont {Senno}, \citenamefont {Murase},\ and\ \citenamefont {Meszaros}}]{Senno:2016bso}%
  \BibitemOpen
  \bibfield  {author} {\bibinfo {author} {\bibfnamefont {N.}~\bibnamefont {Senno}}, \bibinfo {author} {\bibfnamefont {K.}~\bibnamefont {Murase}}, \ and\ \bibinfo {author} {\bibfnamefont {P.}~\bibnamefont {Meszaros}},\ }\href {\doibase 10.3847/1538-4357/aa6344} {\bibfield  {journal} {\bibinfo  {journal} {Astrophys. J.}\ }\textbf {\bibinfo {volume} {838}},\ \bibinfo {pages} {3} (\bibinfo {year} {2017})},\ \Eprint {http://arxiv.org/abs/1612.00918} {arXiv:1612.00918 [astro-ph.HE]} \BibitemShut {NoStop}%
\bibitem [{\citenamefont {Biehl}\ \emph {et~al.}(2018)\citenamefont {Biehl}, \citenamefont {Boncioli}, \citenamefont {Lunardini},\ and\ \citenamefont {Winter}}]{Biehl:2017hnb}%
  \BibitemOpen
  \bibfield  {author} {\bibinfo {author} {\bibfnamefont {D.}~\bibnamefont {Biehl}}, \bibinfo {author} {\bibfnamefont {D.}~\bibnamefont {Boncioli}}, \bibinfo {author} {\bibfnamefont {C.}~\bibnamefont {Lunardini}}, \ and\ \bibinfo {author} {\bibfnamefont {W.}~\bibnamefont {Winter}},\ }\href {\doibase 10.1038/s41598-018-29022-4} {\bibfield  {journal} {\bibinfo  {journal} {Sci. Rep.}\ }\textbf {\bibinfo {volume} {8}},\ \bibinfo {pages} {10828} (\bibinfo {year} {2018})},\ \Eprint {http://arxiv.org/abs/1711.03555} {arXiv:1711.03555 [astro-ph.HE]} \BibitemShut {NoStop}%
\bibitem [{\citenamefont {Zhang}\ \emph {et~al.}(2017)\citenamefont {Zhang}, \citenamefont {Murase}, \citenamefont {Oikonomou},\ and\ \citenamefont {Li}}]{Zhang:2017hom}%
  \BibitemOpen
  \bibfield  {author} {\bibinfo {author} {\bibfnamefont {B.~T.}\ \bibnamefont {Zhang}}, \bibinfo {author} {\bibfnamefont {K.}~\bibnamefont {Murase}}, \bibinfo {author} {\bibfnamefont {F.}~\bibnamefont {Oikonomou}}, \ and\ \bibinfo {author} {\bibfnamefont {Z.}~\bibnamefont {Li}},\ }\href {\doibase 10.1103/PhysRevD.96.063007} {\bibfield  {journal} {\bibinfo  {journal} {Phys. Rev. D}\ }\textbf {\bibinfo {volume} {96}},\ \bibinfo {pages} {063007} (\bibinfo {year} {2017})},\ \bibinfo {note} {[Addendum: Phys.Rev.D 96, 069902 (2017)]},\ \Eprint {http://arxiv.org/abs/1706.00391} {arXiv:1706.00391 [astro-ph.HE]} \BibitemShut {NoStop}%
\bibitem [{\citenamefont {Gu\'epin}\ \emph {et~al.}(2018)\citenamefont {Gu\'epin}, \citenamefont {Kotera}, \citenamefont {Barausse}, \citenamefont {Fang},\ and\ \citenamefont {Murase}}]{Guepin:2017abw}%
  \BibitemOpen
  \bibfield  {author} {\bibinfo {author} {\bibfnamefont {C.}~\bibnamefont {Gu\'epin}}, \bibinfo {author} {\bibfnamefont {K.}~\bibnamefont {Kotera}}, \bibinfo {author} {\bibfnamefont {E.}~\bibnamefont {Barausse}}, \bibinfo {author} {\bibfnamefont {K.}~\bibnamefont {Fang}}, \ and\ \bibinfo {author} {\bibfnamefont {K.}~\bibnamefont {Murase}},\ }\href {\doibase 10.1051/0004-6361/201732392} {\bibfield  {journal} {\bibinfo  {journal} {Astron. Astrophys.}\ }\textbf {\bibinfo {volume} {616}},\ \bibinfo {pages} {A179} (\bibinfo {year} {2018})},\ \bibinfo {note} {[Erratum: Astron.Astrophys. 636, C3 (2020)]},\ \Eprint {http://arxiv.org/abs/1711.11274} {arXiv:1711.11274 [astro-ph.HE]} \BibitemShut {NoStop}%
\bibitem [{\citenamefont {Murase}\ \emph {et~al.}(2020)\citenamefont {Murase}, \citenamefont {Kimura}, \citenamefont {Zhang}, \citenamefont {Oikonomou},\ and\ \citenamefont {Petropoulou}}]{Murase:2020lnu}%
  \BibitemOpen
  \bibfield  {author} {\bibinfo {author} {\bibfnamefont {K.}~\bibnamefont {Murase}}, \bibinfo {author} {\bibfnamefont {S.~S.}\ \bibnamefont {Kimura}}, \bibinfo {author} {\bibfnamefont {B.~T.}\ \bibnamefont {Zhang}}, \bibinfo {author} {\bibfnamefont {F.}~\bibnamefont {Oikonomou}}, \ and\ \bibinfo {author} {\bibfnamefont {M.}~\bibnamefont {Petropoulou}},\ }\href {\doibase 10.3847/1538-4357/abb3c0} {\bibfield  {journal} {\bibinfo  {journal} {Astrophys. J.}\ }\textbf {\bibinfo {volume} {902}},\ \bibinfo {pages} {108} (\bibinfo {year} {2020})},\ \Eprint {http://arxiv.org/abs/2005.08937} {arXiv:2005.08937 [astro-ph.HE]} \BibitemShut {NoStop}%
\bibitem [{\citenamefont {Chan}\ \emph {et~al.}(2021)\citenamefont {Chan}, \citenamefont {Piran},\ and\ \citenamefont {Krolik}}]{Chan:2021blg}%
  \BibitemOpen
  \bibfield  {author} {\bibinfo {author} {\bibfnamefont {C.-H.}\ \bibnamefont {Chan}}, \bibinfo {author} {\bibfnamefont {T.}~\bibnamefont {Piran}}, \ and\ \bibinfo {author} {\bibfnamefont {J.~H.}\ \bibnamefont {Krolik}},\ }\href {\doibase 10.3847/1538-4357/abf0a7} {\bibfield  {journal} {\bibinfo  {journal} {Astrophys. J.}\ }\textbf {\bibinfo {volume} {914}},\ \bibinfo {pages} {107} (\bibinfo {year} {2021})},\ \Eprint {http://arxiv.org/abs/2101.02290} {arXiv:2101.02290 [astro-ph.HE]} \BibitemShut {NoStop}%
\bibitem [{\citenamefont {Mukhopadhyay}\ \emph {et~al.}(2023)\citenamefont {Mukhopadhyay}, \citenamefont {Bhattacharya},\ and\ \citenamefont {Murase}}]{Mukhopadhyay:2023mld}%
  \BibitemOpen
  \bibfield  {author} {\bibinfo {author} {\bibfnamefont {M.}~\bibnamefont {Mukhopadhyay}}, \bibinfo {author} {\bibfnamefont {M.}~\bibnamefont {Bhattacharya}}, \ and\ \bibinfo {author} {\bibfnamefont {K.}~\bibnamefont {Murase}},\ }\href@noop {} {\  (\bibinfo {year} {2023})},\ \Eprint {http://arxiv.org/abs/2309.02275} {arXiv:2309.02275 [astro-ph.HE]} \BibitemShut {NoStop}%
\bibitem [{\citenamefont {Yuan}\ and\ \citenamefont {Winter}(2023)}]{Yuan:2023cmd}%
  \BibitemOpen
  \bibfield  {author} {\bibinfo {author} {\bibfnamefont {C.}~\bibnamefont {Yuan}}\ and\ \bibinfo {author} {\bibfnamefont {W.}~\bibnamefont {Winter}},\ }\href {\doibase 10.3847/1538-4357/acf615} {\bibfield  {journal} {\bibinfo  {journal} {Astrophys. J.}\ }\textbf {\bibinfo {volume} {956}},\ \bibinfo {pages} {30} (\bibinfo {year} {2023})},\ \Eprint {http://arxiv.org/abs/2306.15659} {arXiv:2306.15659 [astro-ph.HE]} \BibitemShut {NoStop}%
\bibitem [{\citenamefont {Stein}\ \emph {et~al.}(2021)\citenamefont {Stein} \emph {et~al.}}]{Stein:2020xhk}%
  \BibitemOpen
  \bibfield  {author} {\bibinfo {author} {\bibfnamefont {R.}~\bibnamefont {Stein}} \emph {et~al.},\ }\href {\doibase 10.1038/s41550-020-01295-8} {\bibfield  {journal} {\bibinfo  {journal} {Nature Astron.}\ }\textbf {\bibinfo {volume} {5}},\ \bibinfo {pages} {510} (\bibinfo {year} {2021})},\ \Eprint {http://arxiv.org/abs/2005.05340} {arXiv:2005.05340 [astro-ph.HE]} \BibitemShut {NoStop}%
\bibitem [{\citenamefont {Reusch}\ \emph {et~al.}(2022{\natexlab{a}})\citenamefont {Reusch} \emph {et~al.}}]{Reusch:2021ztx}%
  \BibitemOpen
  \bibfield  {author} {\bibinfo {author} {\bibfnamefont {S.}~\bibnamefont {Reusch}} \emph {et~al.},\ }\href {\doibase 10.1103/PhysRevLett.128.221101} {\bibfield  {journal} {\bibinfo  {journal} {Phys. Rev. Lett.}\ }\textbf {\bibinfo {volume} {128}},\ \bibinfo {pages} {221101} (\bibinfo {year} {2022}{\natexlab{a}})},\ \Eprint {http://arxiv.org/abs/2111.09390} {arXiv:2111.09390 [astro-ph.HE]} \BibitemShut {NoStop}%
\bibitem [{\citenamefont {van Velzen}\ \emph {et~al.}(2021{\natexlab{a}})\citenamefont {van Velzen} \emph {et~al.}}]{vanVelzen:2021zsm}%
  \BibitemOpen
  \bibfield  {author} {\bibinfo {author} {\bibfnamefont {S.}~\bibnamefont {van Velzen}} \emph {et~al.},\ }\href@noop {} {\  (\bibinfo {year} {2021}{\natexlab{a}})},\ \Eprint {http://arxiv.org/abs/2111.09391} {arXiv:2111.09391 [astro-ph.HE]} \BibitemShut {NoStop}%
\bibitem [{\citenamefont {Winter}\ and\ \citenamefont {Lunardini}(2023)}]{Winter:2022fpf}%
  \BibitemOpen
  \bibfield  {author} {\bibinfo {author} {\bibfnamefont {W.}~\bibnamefont {Winter}}\ and\ \bibinfo {author} {\bibfnamefont {C.}~\bibnamefont {Lunardini}},\ }\href {\doibase 10.3847/1538-4357/acbe9e} {\bibfield  {journal} {\bibinfo  {journal} {Astrophys. J.}\ }\textbf {\bibinfo {volume} {948}},\ \bibinfo {pages} {42} (\bibinfo {year} {2023})},\ \Eprint {http://arxiv.org/abs/2205.11538} {arXiv:2205.11538 [astro-ph.HE]} \BibitemShut {NoStop}%
\bibitem [{\citenamefont {Aartsen}\ \emph {et~al.}(2018)\citenamefont {Aartsen} \emph {et~al.}}]{IceCube:2018dnn}%
  \BibitemOpen
  \bibfield  {author} {\bibinfo {author} {\bibfnamefont {M.~G.}\ \bibnamefont {Aartsen}} \emph {et~al.} (\bibinfo {collaboration} {IceCube, Fermi-LAT, MAGIC, AGILE, ASAS-SN, HAWC, H.E.S.S., INTEGRAL, Kanata, Kiso, Kapteyn, Liverpool Telescope, Subaru, Swift NuSTAR, VERITAS, VLA/17B-403}),\ }\href {\doibase 10.1126/science.aat1378} {\bibfield  {journal} {\bibinfo  {journal} {Science}\ }\textbf {\bibinfo {volume} {361}},\ \bibinfo {pages} {eaat1378} (\bibinfo {year} {2018})},\ \Eprint {http://arxiv.org/abs/1807.08816} {arXiv:1807.08816 [astro-ph.HE]} \BibitemShut {NoStop}%
\bibitem [{\citenamefont {Abbasi}\ \emph {et~al.}(2022)\citenamefont {Abbasi} \emph {et~al.}}]{IceCube:2022der}%
  \BibitemOpen
  \bibfield  {author} {\bibinfo {author} {\bibfnamefont {R.}~\bibnamefont {Abbasi}} \emph {et~al.} (\bibinfo {collaboration} {IceCube}),\ }\href {\doibase 10.1126/science.abg3395} {\bibfield  {journal} {\bibinfo  {journal} {Science}\ }\textbf {\bibinfo {volume} {378}},\ \bibinfo {pages} {538} (\bibinfo {year} {2022})},\ \Eprint {http://arxiv.org/abs/2211.09972} {arXiv:2211.09972 [astro-ph.HE]} \BibitemShut {NoStop}%
\bibitem [{\citenamefont {Ferrer}\ \emph {et~al.}(2023)\citenamefont {Ferrer}, \citenamefont {Herrera},\ and\ \citenamefont {Ibarra}}]{Ferrer:2022kei}%
  \BibitemOpen
  \bibfield  {author} {\bibinfo {author} {\bibfnamefont {F.}~\bibnamefont {Ferrer}}, \bibinfo {author} {\bibfnamefont {G.}~\bibnamefont {Herrera}}, \ and\ \bibinfo {author} {\bibfnamefont {A.}~\bibnamefont {Ibarra}},\ }\href {\doibase 10.1088/1475-7516/2023/05/057} {\bibfield  {journal} {\bibinfo  {journal} {JCAP}\ }\textbf {\bibinfo {volume} {05}},\ \bibinfo {pages} {057} (\bibinfo {year} {2023})},\ \Eprint {http://arxiv.org/abs/2209.06339} {arXiv:2209.06339 [hep-ph]} \BibitemShut {NoStop}%
\bibitem [{\citenamefont {Cline}\ \emph {et~al.}(2023)\citenamefont {Cline}, \citenamefont {Gao}, \citenamefont {Guo}, \citenamefont {Lin}, \citenamefont {Liu}, \citenamefont {Puel}, \citenamefont {Todd},\ and\ \citenamefont {Xiao}}]{Cline:2022qld}%
  \BibitemOpen
  \bibfield  {author} {\bibinfo {author} {\bibfnamefont {J.~M.}\ \bibnamefont {Cline}}, \bibinfo {author} {\bibfnamefont {S.}~\bibnamefont {Gao}}, \bibinfo {author} {\bibfnamefont {F.}~\bibnamefont {Guo}}, \bibinfo {author} {\bibfnamefont {Z.}~\bibnamefont {Lin}}, \bibinfo {author} {\bibfnamefont {S.}~\bibnamefont {Liu}}, \bibinfo {author} {\bibfnamefont {M.}~\bibnamefont {Puel}}, \bibinfo {author} {\bibfnamefont {P.}~\bibnamefont {Todd}}, \ and\ \bibinfo {author} {\bibfnamefont {T.}~\bibnamefont {Xiao}},\ }\href {\doibase 10.1103/PhysRevLett.130.091402} {\bibfield  {journal} {\bibinfo  {journal} {Phys. Rev. Lett.}\ }\textbf {\bibinfo {volume} {130}},\ \bibinfo {pages} {091402} (\bibinfo {year} {2023})},\ \Eprint {http://arxiv.org/abs/2209.02713} {arXiv:2209.02713 [hep-ph]} \BibitemShut {NoStop}%
\bibitem [{\citenamefont {Cline}\ and\ \citenamefont {Puel}(2023)}]{Cline:2023tkp}%
  \BibitemOpen
  \bibfield  {author} {\bibinfo {author} {\bibfnamefont {J.~M.}\ \bibnamefont {Cline}}\ and\ \bibinfo {author} {\bibfnamefont {M.}~\bibnamefont {Puel}},\ }\href@noop {} {\  (\bibinfo {year} {2023})},\ \Eprint {http://arxiv.org/abs/2301.08756} {arXiv:2301.08756 [hep-ph]} \BibitemShut {NoStop}%
\bibitem [{\citenamefont {Kang}\ \emph {et~al.}(2023)\citenamefont {Kang} \emph {et~al.}}]{IceCube:2023cwx}%
  \BibitemOpen
  \bibfield  {author} {\bibinfo {author} {\bibfnamefont {W.}~\bibnamefont {Kang}} \emph {et~al.} (\bibinfo {collaboration} {IceCube}),\ }\href {\doibase 10.22323/1.444.1380} {\bibfield  {journal} {\bibinfo  {journal} {PoS}\ }\textbf {\bibinfo {volume} {ICRC2023}},\ \bibinfo {pages} {1380} (\bibinfo {year} {2023})},\ \Eprint {http://arxiv.org/abs/2308.02842} {arXiv:2308.02842 [astro-ph.HE]} \BibitemShut {NoStop}%
\bibitem [{\citenamefont {Herrera}\ and\ \citenamefont {Murase}(2023)}]{Herrera:2023nww}%
  \BibitemOpen
  \bibfield  {author} {\bibinfo {author} {\bibfnamefont {G.}~\bibnamefont {Herrera}}\ and\ \bibinfo {author} {\bibfnamefont {K.}~\bibnamefont {Murase}},\ }\href@noop {} {\  (\bibinfo {year} {2023})},\ \Eprint {http://arxiv.org/abs/2307.09460} {arXiv:2307.09460 [hep-ph]} \BibitemShut {NoStop}%
\bibitem [{\citenamefont {Choi}\ \emph {et~al.}(2019)\citenamefont {Choi}, \citenamefont {Kim},\ and\ \citenamefont {Rott}}]{Choi:2019ixb}%
  \BibitemOpen
  \bibfield  {author} {\bibinfo {author} {\bibfnamefont {K.-Y.}\ \bibnamefont {Choi}}, \bibinfo {author} {\bibfnamefont {J.}~\bibnamefont {Kim}}, \ and\ \bibinfo {author} {\bibfnamefont {C.}~\bibnamefont {Rott}},\ }\href {\doibase 10.1103/PhysRevD.99.083018} {\bibfield  {journal} {\bibinfo  {journal} {Phys. Rev. D}\ }\textbf {\bibinfo {volume} {99}},\ \bibinfo {pages} {083018} (\bibinfo {year} {2019})},\ \Eprint {http://arxiv.org/abs/1903.03302} {arXiv:1903.03302 [astro-ph.CO]} \BibitemShut {NoStop}%
\bibitem [{\citenamefont {Kelly}\ and\ \citenamefont {Machado}(2018)}]{Kelly:2018tyg}%
  \BibitemOpen
  \bibfield  {author} {\bibinfo {author} {\bibfnamefont {K.~J.}\ \bibnamefont {Kelly}}\ and\ \bibinfo {author} {\bibfnamefont {P.~A.~N.}\ \bibnamefont {Machado}},\ }\href {\doibase 10.1088/1475-7516/2018/10/048} {\bibfield  {journal} {\bibinfo  {journal} {JCAP}\ }\textbf {\bibinfo {volume} {10}},\ \bibinfo {pages} {048} (\bibinfo {year} {2018})},\ \Eprint {http://arxiv.org/abs/1808.02889} {arXiv:1808.02889 [hep-ph]} \BibitemShut {NoStop}%
\bibitem [{\citenamefont {Alvey}\ and\ \citenamefont {Fairbairn}(2019)}]{Alvey_2019}%
  \BibitemOpen
  \bibfield  {author} {\bibinfo {author} {\bibfnamefont {J.}~\bibnamefont {Alvey}}\ and\ \bibinfo {author} {\bibfnamefont {M.}~\bibnamefont {Fairbairn}},\ }\href {\doibase 10.1088/1475-7516/2019/07/041} {\bibfield  {journal} {\bibinfo  {journal} {Journal of Cosmology and Astroparticle Physics}\ }\textbf {\bibinfo {volume} {2019}},\ \bibinfo {pages} {041} (\bibinfo {year} {2019})}\BibitemShut {NoStop}%
\bibitem [{\citenamefont {Argüelles}\ \emph {et~al.}(2017)\citenamefont {Argüelles}, \citenamefont {Kheirandish},\ and\ \citenamefont {Vincent}}]{Arg_elles_2017}%
  \BibitemOpen
  \bibfield  {author} {\bibinfo {author} {\bibfnamefont {C.~A.}\ \bibnamefont {Argüelles}}, \bibinfo {author} {\bibfnamefont {A.}~\bibnamefont {Kheirandish}}, \ and\ \bibinfo {author} {\bibfnamefont {A.~C.}\ \bibnamefont {Vincent}},\ }\href {\doibase 10.1103/physrevlett.119.201801} {\bibfield  {journal} {\bibinfo  {journal} {Physical Review Letters}\ }\textbf {\bibinfo {volume} {119}} (\bibinfo {year} {2017}),\ 10.1103/physrevlett.119.201801}\BibitemShut {NoStop}%
\bibitem [{\citenamefont {Murase}\ and\ \citenamefont {Shoemaker}(2019)}]{Murase:2019xqi}%
  \BibitemOpen
  \bibfield  {author} {\bibinfo {author} {\bibfnamefont {K.}~\bibnamefont {Murase}}\ and\ \bibinfo {author} {\bibfnamefont {I.~M.}\ \bibnamefont {Shoemaker}},\ }\href {\doibase 10.1103/PhysRevLett.123.241102} {\bibfield  {journal} {\bibinfo  {journal} {Phys. Rev. Lett.}\ }\textbf {\bibinfo {volume} {123}},\ \bibinfo {pages} {241102} (\bibinfo {year} {2019})},\ \Eprint {http://arxiv.org/abs/1903.08607} {arXiv:1903.08607 [hep-ph]} \BibitemShut {NoStop}%
\bibitem [{\citenamefont {Brax}\ \emph {et~al.}(2023{\natexlab{a}})\citenamefont {Brax}, \citenamefont {van~de Bruck}, \citenamefont {Di~Valentino}, \citenamefont {Giar\`e},\ and\ \citenamefont {Trojanowski}}]{Brax:2023tvn}%
  \BibitemOpen
  \bibfield  {author} {\bibinfo {author} {\bibfnamefont {P.}~\bibnamefont {Brax}}, \bibinfo {author} {\bibfnamefont {C.}~\bibnamefont {van~de Bruck}}, \bibinfo {author} {\bibfnamefont {E.}~\bibnamefont {Di~Valentino}}, \bibinfo {author} {\bibfnamefont {W.}~\bibnamefont {Giar\`e}}, \ and\ \bibinfo {author} {\bibfnamefont {S.}~\bibnamefont {Trojanowski}},\ }\href {\doibase 10.1016/j.dark.2023.101321} {\bibfield  {journal} {\bibinfo  {journal} {Phys. Dark Univ.}\ }\textbf {\bibinfo {volume} {42}},\ \bibinfo {pages} {101321} (\bibinfo {year} {2023}{\natexlab{a}})},\ \Eprint {http://arxiv.org/abs/2305.01383} {arXiv:2305.01383 [astro-ph.CO]} \BibitemShut {NoStop}%
\bibitem [{\citenamefont {Carpio}\ \emph {et~al.}(2023)\citenamefont {Carpio}, \citenamefont {Kheirandish},\ and\ \citenamefont {Murase}}]{Carpio:2022sml}%
  \BibitemOpen
  \bibfield  {author} {\bibinfo {author} {\bibfnamefont {J.~A.}\ \bibnamefont {Carpio}}, \bibinfo {author} {\bibfnamefont {A.}~\bibnamefont {Kheirandish}}, \ and\ \bibinfo {author} {\bibfnamefont {K.}~\bibnamefont {Murase}},\ }\href {\doibase 10.1088/1475-7516/2023/04/019} {\bibfield  {journal} {\bibinfo  {journal} {JCAP}\ }\textbf {\bibinfo {volume} {04}},\ \bibinfo {pages} {019} (\bibinfo {year} {2023})},\ \Eprint {http://arxiv.org/abs/2204.09650} {arXiv:2204.09650 [hep-ph]} \BibitemShut {NoStop}%
\bibitem [{\citenamefont {A.}\ \emph {et~al.}(2023)\citenamefont {A.}, \citenamefont {Das}, \citenamefont {Lambiase}, \citenamefont {Nomura},\ and\ \citenamefont {Orikasa}}]{A:2023wup}%
  \BibitemOpen
  \bibfield  {author} {\bibinfo {author} {\bibfnamefont {S.~K.}\ \bibnamefont {A.}}, \bibinfo {author} {\bibfnamefont {A.}~\bibnamefont {Das}}, \bibinfo {author} {\bibfnamefont {G.}~\bibnamefont {Lambiase}}, \bibinfo {author} {\bibfnamefont {T.}~\bibnamefont {Nomura}}, \ and\ \bibinfo {author} {\bibfnamefont {Y.}~\bibnamefont {Orikasa}},\ }\href@noop {} {\  (\bibinfo {year} {2023})},\ \Eprint {http://arxiv.org/abs/2308.14483} {arXiv:2308.14483 [hep-ph]} \BibitemShut {NoStop}%
\bibitem [{\citenamefont {Lin}\ \emph {et~al.}(2023)\citenamefont {Lin}, \citenamefont {Tsai}, \citenamefont {Lin}, \citenamefont {Wong},\ and\ \citenamefont {Wu}}]{Lin:2023nsm}%
  \BibitemOpen
  \bibfield  {author} {\bibinfo {author} {\bibfnamefont {Y.-H.}\ \bibnamefont {Lin}}, \bibinfo {author} {\bibfnamefont {T.-H.}\ \bibnamefont {Tsai}}, \bibinfo {author} {\bibfnamefont {G.-L.}\ \bibnamefont {Lin}}, \bibinfo {author} {\bibfnamefont {H.~T.-K.}\ \bibnamefont {Wong}}, \ and\ \bibinfo {author} {\bibfnamefont {M.-R.}\ \bibnamefont {Wu}},\ }\href {\doibase 10.1103/PhysRevD.108.083013} {\bibfield  {journal} {\bibinfo  {journal} {Phys. Rev. D}\ }\textbf {\bibinfo {volume} {108}},\ \bibinfo {pages} {083013} (\bibinfo {year} {2023})},\ \Eprint {http://arxiv.org/abs/2307.03522} {arXiv:2307.03522 [hep-ph]} \BibitemShut {NoStop}%
\bibitem [{\citenamefont {Koren}(2019)}]{Koren:2019wwi}%
  \BibitemOpen
  \bibfield  {author} {\bibinfo {author} {\bibfnamefont {S.}~\bibnamefont {Koren}},\ }\href {\doibase 10.1088/1475-7516/2019/09/013} {\bibfield  {journal} {\bibinfo  {journal} {JCAP}\ }\textbf {\bibinfo {volume} {09}},\ \bibinfo {pages} {013} (\bibinfo {year} {2019})},\ \Eprint {http://arxiv.org/abs/1903.05096} {arXiv:1903.05096 [hep-ph]} \BibitemShut {NoStop}%
\bibitem [{\citenamefont {McMullen}\ \emph {et~al.}(2021)\citenamefont {McMullen}, \citenamefont {Vincent}, \citenamefont {Arguelles},\ and\ \citenamefont {Schneider}}]{McMullen:2021ikf}%
  \BibitemOpen
  \bibfield  {author} {\bibinfo {author} {\bibfnamefont {A.}~\bibnamefont {McMullen}}, \bibinfo {author} {\bibfnamefont {A.}~\bibnamefont {Vincent}}, \bibinfo {author} {\bibfnamefont {C.}~\bibnamefont {Arguelles}}, \ and\ \bibinfo {author} {\bibfnamefont {A.}~\bibnamefont {Schneider}} (\bibinfo {collaboration} {IceCube}),\ }\href {\doibase 10.1088/1748-0221/16/08/C08001} {\bibfield  {journal} {\bibinfo  {journal} {JINST}\ }\textbf {\bibinfo {volume} {16}},\ \bibinfo {pages} {C08001} (\bibinfo {year} {2021})},\ \Eprint {http://arxiv.org/abs/2107.11491} {arXiv:2107.11491 [astro-ph.HE]} \BibitemShut {NoStop}%
\bibitem [{\citenamefont {Mangano}\ \emph {et~al.}(2006)\citenamefont {Mangano}, \citenamefont {Melchiorri}, \citenamefont {Serra}, \citenamefont {Cooray},\ and\ \citenamefont {Kamionkowski}}]{Mangano_2006}%
  \BibitemOpen
  \bibfield  {author} {\bibinfo {author} {\bibfnamefont {G.}~\bibnamefont {Mangano}}, \bibinfo {author} {\bibfnamefont {A.}~\bibnamefont {Melchiorri}}, \bibinfo {author} {\bibfnamefont {P.}~\bibnamefont {Serra}}, \bibinfo {author} {\bibfnamefont {A.}~\bibnamefont {Cooray}}, \ and\ \bibinfo {author} {\bibfnamefont {M.}~\bibnamefont {Kamionkowski}},\ }\href {\doibase 10.1103/physrevd.74.043517} {\bibfield  {journal} {\bibinfo  {journal} {Physical Review D}\ }\textbf {\bibinfo {volume} {74}} (\bibinfo {year} {2006}),\ 10.1103/physrevd.74.043517}\BibitemShut {NoStop}%
\bibitem [{\citenamefont {Wilkinson}\ \emph {et~al.}(2014{\natexlab{a}})\citenamefont {Wilkinson}, \citenamefont {Lesgourgues},\ and\ \citenamefont {Boehm}}]{Wilkinson:2013kia}%
  \BibitemOpen
  \bibfield  {author} {\bibinfo {author} {\bibfnamefont {R.~J.}\ \bibnamefont {Wilkinson}}, \bibinfo {author} {\bibfnamefont {J.}~\bibnamefont {Lesgourgues}}, \ and\ \bibinfo {author} {\bibfnamefont {C.}~\bibnamefont {Boehm}},\ }\href {\doibase 10.1088/1475-7516/2014/04/026} {\bibfield  {journal} {\bibinfo  {journal} {JCAP}\ }\textbf {\bibinfo {volume} {04}},\ \bibinfo {pages} {026} (\bibinfo {year} {2014}{\natexlab{a}})},\ \Eprint {http://arxiv.org/abs/1309.7588} {arXiv:1309.7588 [astro-ph.CO]} \BibitemShut {NoStop}%
\bibitem [{\citenamefont {Wilkinson}\ \emph {et~al.}(2014{\natexlab{b}})\citenamefont {Wilkinson}, \citenamefont {Boehm},\ and\ \citenamefont {Lesgourgues}}]{Wilkinson:2014ksa}%
  \BibitemOpen
  \bibfield  {author} {\bibinfo {author} {\bibfnamefont {R.~J.}\ \bibnamefont {Wilkinson}}, \bibinfo {author} {\bibfnamefont {C.}~\bibnamefont {Boehm}}, \ and\ \bibinfo {author} {\bibfnamefont {J.}~\bibnamefont {Lesgourgues}},\ }\href {\doibase 10.1088/1475-7516/2014/05/011} {\bibfield  {journal} {\bibinfo  {journal} {JCAP}\ }\textbf {\bibinfo {volume} {05}},\ \bibinfo {pages} {011} (\bibinfo {year} {2014}{\natexlab{b}})},\ \Eprint {http://arxiv.org/abs/1401.7597} {arXiv:1401.7597 [astro-ph.CO]} \BibitemShut {NoStop}%
\bibitem [{\citenamefont {Escudero}\ \emph {et~al.}(2015)\citenamefont {Escudero}, \citenamefont {Mena}, \citenamefont {Vincent}, \citenamefont {Wilkinson},\ and\ \citenamefont {B\oe{}hm}}]{Escudero:2015yka}%
  \BibitemOpen
  \bibfield  {author} {\bibinfo {author} {\bibfnamefont {M.}~\bibnamefont {Escudero}}, \bibinfo {author} {\bibfnamefont {O.}~\bibnamefont {Mena}}, \bibinfo {author} {\bibfnamefont {A.~C.}\ \bibnamefont {Vincent}}, \bibinfo {author} {\bibfnamefont {R.~J.}\ \bibnamefont {Wilkinson}}, \ and\ \bibinfo {author} {\bibfnamefont {C.}~\bibnamefont {B\oe{}hm}},\ }\href {\doibase 10.1088/1475-7516/2015/9/034} {\bibfield  {journal} {\bibinfo  {journal} {JCAP}\ }\textbf {\bibinfo {volume} {09}},\ \bibinfo {pages} {034} (\bibinfo {year} {2015})},\ \Eprint {http://arxiv.org/abs/1505.06735} {arXiv:1505.06735 [astro-ph.CO]} \BibitemShut {NoStop}%
\bibitem [{\citenamefont {Essig}\ \emph {et~al.}(2017)\citenamefont {Essig}, \citenamefont {Volansky},\ and\ \citenamefont {Yu}}]{Essig:2017kqs}%
  \BibitemOpen
  \bibfield  {author} {\bibinfo {author} {\bibfnamefont {R.}~\bibnamefont {Essig}}, \bibinfo {author} {\bibfnamefont {T.}~\bibnamefont {Volansky}}, \ and\ \bibinfo {author} {\bibfnamefont {T.-T.}\ \bibnamefont {Yu}},\ }\href {\doibase 10.1103/PhysRevD.96.043017} {\bibfield  {journal} {\bibinfo  {journal} {Phys. Rev. D}\ }\textbf {\bibinfo {volume} {96}},\ \bibinfo {pages} {043017} (\bibinfo {year} {2017})},\ \Eprint {http://arxiv.org/abs/1703.00910} {arXiv:1703.00910 [hep-ph]} \BibitemShut {NoStop}%
\bibitem [{\citenamefont {Aprile}\ \emph {et~al.}(2019)\citenamefont {Aprile} \emph {et~al.}}]{XENON:2019gfn}%
  \BibitemOpen
  \bibfield  {author} {\bibinfo {author} {\bibfnamefont {E.}~\bibnamefont {Aprile}} \emph {et~al.} (\bibinfo {collaboration} {XENON}),\ }\href {\doibase 10.1103/PhysRevLett.123.251801} {\bibfield  {journal} {\bibinfo  {journal} {Phys. Rev. Lett.}\ }\textbf {\bibinfo {volume} {123}},\ \bibinfo {pages} {251801} (\bibinfo {year} {2019})},\ \Eprint {http://arxiv.org/abs/1907.11485} {arXiv:1907.11485 [hep-ex]} \BibitemShut {NoStop}%
\bibitem [{\citenamefont {Jho}\ \emph {et~al.}(2021)\citenamefont {Jho}, \citenamefont {Park}, \citenamefont {Park},\ and\ \citenamefont {Tseng}}]{Jho:2021rmn}%
  \BibitemOpen
  \bibfield  {author} {\bibinfo {author} {\bibfnamefont {Y.}~\bibnamefont {Jho}}, \bibinfo {author} {\bibfnamefont {J.-C.}\ \bibnamefont {Park}}, \bibinfo {author} {\bibfnamefont {S.~C.}\ \bibnamefont {Park}}, \ and\ \bibinfo {author} {\bibfnamefont {P.-Y.}\ \bibnamefont {Tseng}},\ }\href@noop {} {\  (\bibinfo {year} {2021})},\ \Eprint {http://arxiv.org/abs/2101.11262} {arXiv:2101.11262 [hep-ph]} \BibitemShut {NoStop}%
\bibitem [{\citenamefont {Zhang}(2022)}]{Zhang:2020nis}%
  \BibitemOpen
  \bibfield  {author} {\bibinfo {author} {\bibfnamefont {Y.}~\bibnamefont {Zhang}},\ }\href {\doibase 10.1093/ptep/ptab156} {\bibfield  {journal} {\bibinfo  {journal} {PTEP}\ }\textbf {\bibinfo {volume} {2022}},\ \bibinfo {pages} {013B05} (\bibinfo {year} {2022})},\ \Eprint {http://arxiv.org/abs/2001.00948} {arXiv:2001.00948 [hep-ph]} \BibitemShut {NoStop}%
\bibitem [{\citenamefont {Ghosh}\ \emph {et~al.}(2022)\citenamefont {Ghosh}, \citenamefont {Guha},\ and\ \citenamefont {Sachdeva}}]{Ghosh:2021vkt}%
  \BibitemOpen
  \bibfield  {author} {\bibinfo {author} {\bibfnamefont {D.}~\bibnamefont {Ghosh}}, \bibinfo {author} {\bibfnamefont {A.}~\bibnamefont {Guha}}, \ and\ \bibinfo {author} {\bibfnamefont {D.}~\bibnamefont {Sachdeva}},\ }\href {\doibase 10.1103/PhysRevD.105.103029} {\bibfield  {journal} {\bibinfo  {journal} {Phys. Rev. D}\ }\textbf {\bibinfo {volume} {105}},\ \bibinfo {pages} {103029} (\bibinfo {year} {2022})},\ \Eprint {http://arxiv.org/abs/2110.00025} {arXiv:2110.00025 [hep-ph]} \BibitemShut {NoStop}%
\bibitem [{\citenamefont {Farzan}\ and\ \citenamefont {Palomares-Ruiz}(2014)}]{Farzan:2014gza}%
  \BibitemOpen
  \bibfield  {author} {\bibinfo {author} {\bibfnamefont {Y.}~\bibnamefont {Farzan}}\ and\ \bibinfo {author} {\bibfnamefont {S.}~\bibnamefont {Palomares-Ruiz}},\ }\href {\doibase 10.1088/1475-7516/2014/06/014} {\bibfield  {journal} {\bibinfo  {journal} {JCAP}\ }\textbf {\bibinfo {volume} {06}},\ \bibinfo {pages} {014} (\bibinfo {year} {2014})},\ \Eprint {http://arxiv.org/abs/1401.7019} {arXiv:1401.7019 [hep-ph]} \BibitemShut {NoStop}%
\bibitem [{\citenamefont {Das}\ and\ \citenamefont {Sen}(2021)}]{Das:2021lcr}%
  \BibitemOpen
  \bibfield  {author} {\bibinfo {author} {\bibfnamefont {A.}~\bibnamefont {Das}}\ and\ \bibinfo {author} {\bibfnamefont {M.}~\bibnamefont {Sen}},\ }\href {\doibase 10.1103/PhysRevD.104.075029} {\bibfield  {journal} {\bibinfo  {journal} {Phys. Rev. D}\ }\textbf {\bibinfo {volume} {104}},\ \bibinfo {pages} {075029} (\bibinfo {year} {2021})},\ \Eprint {http://arxiv.org/abs/2104.00027} {arXiv:2104.00027 [hep-ph]} \BibitemShut {NoStop}%
\bibitem [{\citenamefont {Lin}\ \emph {et~al.}(2022)\citenamefont {Lin}, \citenamefont {Wu}, \citenamefont {Wu},\ and\ \citenamefont {Wong}}]{Lin:2022dbl}%
  \BibitemOpen
  \bibfield  {author} {\bibinfo {author} {\bibfnamefont {Y.-H.}\ \bibnamefont {Lin}}, \bibinfo {author} {\bibfnamefont {W.-H.}\ \bibnamefont {Wu}}, \bibinfo {author} {\bibfnamefont {M.-R.}\ \bibnamefont {Wu}}, \ and\ \bibinfo {author} {\bibfnamefont {H.~T.-K.}\ \bibnamefont {Wong}},\ }\href@noop {} {\  (\bibinfo {year} {2022})},\ \Eprint {http://arxiv.org/abs/2206.06864} {arXiv:2206.06864 [hep-ph]} \BibitemShut {NoStop}%
\bibitem [{\citenamefont {Wang}\ \emph {et~al.}(2021)\citenamefont {Wang}, \citenamefont {Granelli},\ and\ \citenamefont {Ullio}}]{Wang:2021jic}%
  \BibitemOpen
  \bibfield  {author} {\bibinfo {author} {\bibfnamefont {J.-W.}\ \bibnamefont {Wang}}, \bibinfo {author} {\bibfnamefont {A.}~\bibnamefont {Granelli}}, \ and\ \bibinfo {author} {\bibfnamefont {P.}~\bibnamefont {Ullio}},\ }\href@noop {} {\  (\bibinfo {year} {2021})},\ \Eprint {http://arxiv.org/abs/2111.13644} {arXiv:2111.13644 [astro-ph.HE]} \BibitemShut {NoStop}%
\bibitem [{\citenamefont {Abe}\ \emph {et~al.}(2023)\citenamefont {Abe} \emph {et~al.}}]{Super-Kamiokande:2022ncz}%
  \BibitemOpen
  \bibfield  {author} {\bibinfo {author} {\bibfnamefont {K.}~\bibnamefont {Abe}} \emph {et~al.} (\bibinfo {collaboration} {Super-Kamiokande}),\ }\href {\doibase 10.1103/PhysRevLett.130.031802} {\bibfield  {journal} {\bibinfo  {journal} {Phys. Rev. Lett.}\ }\textbf {\bibinfo {volume} {130}},\ \bibinfo {pages} {031802} (\bibinfo {year} {2023})},\ \bibinfo {note} {[Erratum: Phys.Rev.Lett. 131, 159903 (2023)]},\ \Eprint {http://arxiv.org/abs/2209.14968} {arXiv:2209.14968 [hep-ex]} \BibitemShut {NoStop}%
\bibitem [{\citenamefont {Alvey}\ \emph {et~al.}(2023)\citenamefont {Alvey}, \citenamefont {Bringmann},\ and\ \citenamefont {Kolesova}}]{Alvey:2022pad}%
  \BibitemOpen
  \bibfield  {author} {\bibinfo {author} {\bibfnamefont {J.}~\bibnamefont {Alvey}}, \bibinfo {author} {\bibfnamefont {T.}~\bibnamefont {Bringmann}}, \ and\ \bibinfo {author} {\bibfnamefont {H.}~\bibnamefont {Kolesova}},\ }\href {\doibase 10.1007/JHEP01(2023)123} {\bibfield  {journal} {\bibinfo  {journal} {JHEP}\ }\textbf {\bibinfo {volume} {01}},\ \bibinfo {pages} {123} (\bibinfo {year} {2023})},\ \Eprint {http://arxiv.org/abs/2209.03360} {arXiv:2209.03360 [hep-ph]} \BibitemShut {NoStop}%
\bibitem [{\citenamefont {An}\ \emph {et~al.}(2021)\citenamefont {An}, \citenamefont {Nie}, \citenamefont {Pospelov}, \citenamefont {Pradler},\ and\ \citenamefont {Ritz}}]{An:2021qdl}%
  \BibitemOpen
  \bibfield  {author} {\bibinfo {author} {\bibfnamefont {H.}~\bibnamefont {An}}, \bibinfo {author} {\bibfnamefont {H.}~\bibnamefont {Nie}}, \bibinfo {author} {\bibfnamefont {M.}~\bibnamefont {Pospelov}}, \bibinfo {author} {\bibfnamefont {J.}~\bibnamefont {Pradler}}, \ and\ \bibinfo {author} {\bibfnamefont {A.}~\bibnamefont {Ritz}},\ }\href {\doibase 10.1103/PhysRevD.104.103026} {\bibfield  {journal} {\bibinfo  {journal} {Phys. Rev. D}\ }\textbf {\bibinfo {volume} {104}},\ \bibinfo {pages} {103026} (\bibinfo {year} {2021})},\ \Eprint {http://arxiv.org/abs/2108.10332} {arXiv:2108.10332 [hep-ph]} \BibitemShut {NoStop}%
\bibitem [{\citenamefont {Guo}\ \emph {et~al.}(2020)\citenamefont {Guo}, \citenamefont {Tsai}, \citenamefont {Wu},\ and\ \citenamefont {Yuan}}]{Guo:2020oum}%
  \BibitemOpen
  \bibfield  {author} {\bibinfo {author} {\bibfnamefont {G.}~\bibnamefont {Guo}}, \bibinfo {author} {\bibfnamefont {Y.-L.~S.}\ \bibnamefont {Tsai}}, \bibinfo {author} {\bibfnamefont {M.-R.}\ \bibnamefont {Wu}}, \ and\ \bibinfo {author} {\bibfnamefont {Q.}~\bibnamefont {Yuan}},\ }\href {\doibase 10.1103/PhysRevD.102.103004} {\bibfield  {journal} {\bibinfo  {journal} {Phys. Rev. D}\ }\textbf {\bibinfo {volume} {102}},\ \bibinfo {pages} {103004} (\bibinfo {year} {2020})},\ \Eprint {http://arxiv.org/abs/2008.12137} {arXiv:2008.12137 [astro-ph.HE]} \BibitemShut {NoStop}%
\bibitem [{\citenamefont {Ambrosone}\ \emph {et~al.}(2023)\citenamefont {Ambrosone}, \citenamefont {Chianese}, \citenamefont {Fiorillo}, \citenamefont {Marinelli},\ and\ \citenamefont {Miele}}]{Ambrosone:2022mvk}%
  \BibitemOpen
  \bibfield  {author} {\bibinfo {author} {\bibfnamefont {A.}~\bibnamefont {Ambrosone}}, \bibinfo {author} {\bibfnamefont {M.}~\bibnamefont {Chianese}}, \bibinfo {author} {\bibfnamefont {D.~F.~G.}\ \bibnamefont {Fiorillo}}, \bibinfo {author} {\bibfnamefont {A.}~\bibnamefont {Marinelli}}, \ and\ \bibinfo {author} {\bibfnamefont {G.}~\bibnamefont {Miele}},\ }\href {\doibase 10.1103/PhysRevLett.131.111003} {\bibfield  {journal} {\bibinfo  {journal} {Phys. Rev. Lett.}\ }\textbf {\bibinfo {volume} {131}},\ \bibinfo {pages} {111003} (\bibinfo {year} {2023})},\ \Eprint {http://arxiv.org/abs/2210.05685} {arXiv:2210.05685 [astro-ph.HE]} \BibitemShut {NoStop}%
\bibitem [{\citenamefont {Bell}\ \emph {et~al.}(2023)\citenamefont {Bell}, \citenamefont {Newstead},\ and\ \citenamefont {Shaukat-Ali}}]{Bell:2023sdq}%
  \BibitemOpen
  \bibfield  {author} {\bibinfo {author} {\bibfnamefont {N.~F.}\ \bibnamefont {Bell}}, \bibinfo {author} {\bibfnamefont {J.~L.}\ \bibnamefont {Newstead}}, \ and\ \bibinfo {author} {\bibfnamefont {I.}~\bibnamefont {Shaukat-Ali}},\ }\href@noop {} {\  (\bibinfo {year} {2023})},\ \Eprint {http://arxiv.org/abs/2309.11003} {arXiv:2309.11003 [hep-ph]} \BibitemShut {NoStop}%
\bibitem [{\citenamefont {John}\ \emph {et~al.}(2023)\citenamefont {John}, \citenamefont {Leane},\ and\ \citenamefont {Linden}}]{John:2023knt}%
  \BibitemOpen
  \bibfield  {author} {\bibinfo {author} {\bibfnamefont {I.}~\bibnamefont {John}}, \bibinfo {author} {\bibfnamefont {R.~K.}\ \bibnamefont {Leane}}, \ and\ \bibinfo {author} {\bibfnamefont {T.}~\bibnamefont {Linden}},\ }\href@noop {} {\  (\bibinfo {year} {2023})},\ \Eprint {http://arxiv.org/abs/2311.16228} {arXiv:2311.16228 [astro-ph.HE]} \BibitemShut {NoStop}%
\bibitem [{\citenamefont {Gondolo}\ and\ \citenamefont {Silk}(1999)}]{Gondolo_1999}%
  \BibitemOpen
  \bibfield  {author} {\bibinfo {author} {\bibfnamefont {P.}~\bibnamefont {Gondolo}}\ and\ \bibinfo {author} {\bibfnamefont {J.}~\bibnamefont {Silk}},\ }\href {\doibase 10.1103/physrevlett.83.1719} {\bibfield  {journal} {\bibinfo  {journal} {Physical Review Letters}\ }\textbf {\bibinfo {volume} {83}},\ \bibinfo {pages} {1719} (\bibinfo {year} {1999})}\BibitemShut {NoStop}%
\bibitem [{\citenamefont {Ryu}\ \emph {et~al.}(2020)\citenamefont {Ryu}, \citenamefont {Krolik}, \citenamefont {Piran},\ and\ \citenamefont {Noble}}]{Ryu:2020cqv}%
  \BibitemOpen
  \bibfield  {author} {\bibinfo {author} {\bibfnamefont {T.}~\bibnamefont {Ryu}}, \bibinfo {author} {\bibfnamefont {J.}~\bibnamefont {Krolik}}, \bibinfo {author} {\bibfnamefont {T.}~\bibnamefont {Piran}}, \ and\ \bibinfo {author} {\bibfnamefont {S.~C.}\ \bibnamefont {Noble}},\ }\href {\doibase 10.3847/1538-4357/abb3cf} {\bibfield  {journal} {\bibinfo  {journal} {Astrophys. J.}\ }\textbf {\bibinfo {volume} {904}},\ \bibinfo {pages} {98} (\bibinfo {year} {2020})},\ \Eprint {http://arxiv.org/abs/2001.03501} {arXiv:2001.03501 [astro-ph.HE]} \BibitemShut {NoStop}%
\bibitem [{\citenamefont {Padovani}\ \emph {et~al.}(2019)\citenamefont {Padovani}, \citenamefont {Oikonomou}, \citenamefont {Petropoulou}, \citenamefont {Giommi},\ and\ \citenamefont {Resconi}}]{Padovani_2019}%
  \BibitemOpen
  \bibfield  {author} {\bibinfo {author} {\bibfnamefont {P.}~\bibnamefont {Padovani}}, \bibinfo {author} {\bibfnamefont {F.}~\bibnamefont {Oikonomou}}, \bibinfo {author} {\bibfnamefont {M.}~\bibnamefont {Petropoulou}}, \bibinfo {author} {\bibfnamefont {P.}~\bibnamefont {Giommi}}, \ and\ \bibinfo {author} {\bibfnamefont {E.}~\bibnamefont {Resconi}},\ }\href {\doibase 10.1093/mnrasl/slz011} {\bibfield  {journal} {\bibinfo  {journal} {Monthly Notices of the Royal Astronomical Society: Letters}\ }\textbf {\bibinfo {volume} {484}},\ \bibinfo {pages} {L104} (\bibinfo {year} {2019})}\BibitemShut {NoStop}%
\bibitem [{\citenamefont {Murase}(2022)}]{Murase:2022dog}%
  \BibitemOpen
  \bibfield  {author} {\bibinfo {author} {\bibfnamefont {K.}~\bibnamefont {Murase}},\ }\href {\doibase 10.3847/2041-8213/aca53c} {\bibfield  {journal} {\bibinfo  {journal} {Astrophys. J. Lett.}\ }\textbf {\bibinfo {volume} {941}},\ \bibinfo {pages} {L17} (\bibinfo {year} {2022})},\ \Eprint {http://arxiv.org/abs/2211.04460} {arXiv:2211.04460 [astro-ph.HE]} \BibitemShut {NoStop}%
\bibitem [{\citenamefont {Eichmann}\ \emph {et~al.}(2022)\citenamefont {Eichmann}, \citenamefont {Oikonomou}, \citenamefont {Salvatore}, \citenamefont {Dettmar},\ and\ \citenamefont {Becker~Tjus}}]{Eichmann:2022lxh}%
  \BibitemOpen
  \bibfield  {author} {\bibinfo {author} {\bibfnamefont {B.}~\bibnamefont {Eichmann}}, \bibinfo {author} {\bibfnamefont {F.}~\bibnamefont {Oikonomou}}, \bibinfo {author} {\bibfnamefont {S.}~\bibnamefont {Salvatore}}, \bibinfo {author} {\bibfnamefont {R.-J.}\ \bibnamefont {Dettmar}}, \ and\ \bibinfo {author} {\bibfnamefont {J.}~\bibnamefont {Becker~Tjus}},\ }\href {\doibase 10.3847/1538-4357/ac9588} {\bibfield  {journal} {\bibinfo  {journal} {Astrophys. J.}\ }\textbf {\bibinfo {volume} {939}},\ \bibinfo {pages} {43} (\bibinfo {year} {2022})},\ \Eprint {http://arxiv.org/abs/2207.00102} {arXiv:2207.00102 [astro-ph.HE]} \BibitemShut {NoStop}%
\bibitem [{\citenamefont {Merritt}(2004)}]{Merritt:2003qk}%
  \BibitemOpen
  \bibfield  {author} {\bibinfo {author} {\bibfnamefont {D.}~\bibnamefont {Merritt}},\ }\href {\doibase 10.1103/PhysRevLett.92.201304} {\bibfield  {journal} {\bibinfo  {journal} {Phys. Rev. Lett.}\ }\textbf {\bibinfo {volume} {92}},\ \bibinfo {pages} {201304} (\bibinfo {year} {2004})},\ \Eprint {http://arxiv.org/abs/astro-ph/0311594} {arXiv:astro-ph/0311594} \BibitemShut {NoStop}%
\bibitem [{\citenamefont {Bertone}\ \emph {et~al.}(2005)\citenamefont {Bertone}, \citenamefont {Hooper},\ and\ \citenamefont {Silk}}]{Bertone:2004pz}%
  \BibitemOpen
  \bibfield  {author} {\bibinfo {author} {\bibfnamefont {G.}~\bibnamefont {Bertone}}, \bibinfo {author} {\bibfnamefont {D.}~\bibnamefont {Hooper}}, \ and\ \bibinfo {author} {\bibfnamefont {J.}~\bibnamefont {Silk}},\ }\href {\doibase 10.1016/j.physrep.2004.08.031} {\bibfield  {journal} {\bibinfo  {journal} {Phys. Rept.}\ }\textbf {\bibinfo {volume} {405}},\ \bibinfo {pages} {279} (\bibinfo {year} {2005})},\ \Eprint {http://arxiv.org/abs/hep-ph/0404175} {arXiv:hep-ph/0404175} \BibitemShut {NoStop}%
\bibitem [{\citenamefont {Merritt}\ \emph {et~al.}(2007)\citenamefont {Merritt}, \citenamefont {Harfst},\ and\ \citenamefont {Bertone}}]{Merritt:2006mt}%
  \BibitemOpen
  \bibfield  {author} {\bibinfo {author} {\bibfnamefont {D.}~\bibnamefont {Merritt}}, \bibinfo {author} {\bibfnamefont {S.}~\bibnamefont {Harfst}}, \ and\ \bibinfo {author} {\bibfnamefont {G.}~\bibnamefont {Bertone}},\ }\href {\doibase 10.1103/PhysRevD.75.043517} {\bibfield  {journal} {\bibinfo  {journal} {Phys. Rev. D}\ }\textbf {\bibinfo {volume} {75}},\ \bibinfo {pages} {043517} (\bibinfo {year} {2007})},\ \Eprint {http://arxiv.org/abs/astro-ph/0610425} {arXiv:astro-ph/0610425} \BibitemShut {NoStop}%
\bibitem [{\citenamefont {Pitik}\ \emph {et~al.}(2022)\citenamefont {Pitik}, \citenamefont {Tamborra}, \citenamefont {Angus},\ and\ \citenamefont {Auchettl}}]{Pitik_2022}%
  \BibitemOpen
  \bibfield  {author} {\bibinfo {author} {\bibfnamefont {T.}~\bibnamefont {Pitik}}, \bibinfo {author} {\bibfnamefont {I.}~\bibnamefont {Tamborra}}, \bibinfo {author} {\bibfnamefont {C.~R.}\ \bibnamefont {Angus}}, \ and\ \bibinfo {author} {\bibfnamefont {K.}~\bibnamefont {Auchettl}},\ }\href {\doibase 10.3847/1538-4357/ac5ab1} {\bibfield  {journal} {\bibinfo  {journal} {The Astrophysical Journal}\ }\textbf {\bibinfo {volume} {929}},\ \bibinfo {pages} {163} (\bibinfo {year} {2022})}\BibitemShut {NoStop}%
\bibitem [{\citenamefont {Dai}\ \emph {et~al.}(2018)\citenamefont {Dai}, \citenamefont {McKinney}, \citenamefont {Roth}, \citenamefont {Ramirez-Ruiz},\ and\ \citenamefont {Miller}}]{Dai_2018}%
  \BibitemOpen
  \bibfield  {author} {\bibinfo {author} {\bibfnamefont {L.}~\bibnamefont {Dai}}, \bibinfo {author} {\bibfnamefont {J.~C.}\ \bibnamefont {McKinney}}, \bibinfo {author} {\bibfnamefont {N.}~\bibnamefont {Roth}}, \bibinfo {author} {\bibfnamefont {E.}~\bibnamefont {Ramirez-Ruiz}}, \ and\ \bibinfo {author} {\bibfnamefont {M.~C.}\ \bibnamefont {Miller}},\ }\href {\doibase 10.3847/2041-8213/aab429} {\bibfield  {journal} {\bibinfo  {journal} {The Astrophysical Journal Letters}\ }\textbf {\bibinfo {volume} {859}},\ \bibinfo {pages} {L20} (\bibinfo {year} {2018})}\BibitemShut {NoStop}%
\bibitem [{\citenamefont {Wang}\ \emph {et~al.}(2011{\natexlab{b}})\citenamefont {Wang}, \citenamefont {Liu}, \citenamefont {Dai},\ and\ \citenamefont {Cheng}}]{PhysRevD.84.081301}%
  \BibitemOpen
  \bibfield  {author} {\bibinfo {author} {\bibfnamefont {X.-Y.}\ \bibnamefont {Wang}}, \bibinfo {author} {\bibfnamefont {R.-Y.}\ \bibnamefont {Liu}}, \bibinfo {author} {\bibfnamefont {Z.-G.}\ \bibnamefont {Dai}}, \ and\ \bibinfo {author} {\bibfnamefont {K.~S.}\ \bibnamefont {Cheng}},\ }\href {\doibase 10.1103/PhysRevD.84.081301} {\bibfield  {journal} {\bibinfo  {journal} {Phys. Rev. D}\ }\textbf {\bibinfo {volume} {84}},\ \bibinfo {pages} {081301} (\bibinfo {year} {2011}{\natexlab{b}})}\BibitemShut {NoStop}%
\bibitem [{\citenamefont {Wang}\ and\ \citenamefont {Liu}(2016)}]{PhysRevD.93.083005}%
  \BibitemOpen
  \bibfield  {author} {\bibinfo {author} {\bibfnamefont {X.-Y.}\ \bibnamefont {Wang}}\ and\ \bibinfo {author} {\bibfnamefont {R.-Y.}\ \bibnamefont {Liu}},\ }\href {\doibase 10.1103/PhysRevD.93.083005} {\bibfield  {journal} {\bibinfo  {journal} {Phys. Rev. D}\ }\textbf {\bibinfo {volume} {93}},\ \bibinfo {pages} {083005} (\bibinfo {year} {2016})}\BibitemShut {NoStop}%
\bibitem [{\citenamefont {Wang}\ \emph {et~al.}(2011{\natexlab{c}})\citenamefont {Wang}, \citenamefont {Liu}, \citenamefont {Dai},\ and\ \citenamefont {Cheng}}]{Wang:2011ip}%
  \BibitemOpen
  \bibfield  {author} {\bibinfo {author} {\bibfnamefont {X.-Y.}\ \bibnamefont {Wang}}, \bibinfo {author} {\bibfnamefont {R.-Y.}\ \bibnamefont {Liu}}, \bibinfo {author} {\bibfnamefont {Z.-G.}\ \bibnamefont {Dai}}, \ and\ \bibinfo {author} {\bibfnamefont {K.~S.}\ \bibnamefont {Cheng}},\ }\href {\doibase 10.1103/PhysRevD.84.081301} {\bibfield  {journal} {\bibinfo  {journal} {Phys. Rev. D}\ }\textbf {\bibinfo {volume} {84}},\ \bibinfo {pages} {081301} (\bibinfo {year} {2011}{\natexlab{c}})},\ \Eprint {http://arxiv.org/abs/1106.2426} {arXiv:1106.2426 [astro-ph.HE]} \BibitemShut {NoStop}%
\bibitem [{\citenamefont {Hayasaki}\ and\ \citenamefont {Yamazaki}(2019)}]{Hayasaki:2019kjy}%
  \BibitemOpen
  \bibfield  {author} {\bibinfo {author} {\bibfnamefont {K.}~\bibnamefont {Hayasaki}}\ and\ \bibinfo {author} {\bibfnamefont {R.}~\bibnamefont {Yamazaki}},\ }\href {\doibase 10.3847/1538-4357/ab44ca} {\  (\bibinfo {year} {2019}),\ 10.3847/1538-4357/ab44ca},\ \Eprint {http://arxiv.org/abs/1908.10882} {arXiv:1908.10882 [astro-ph.HE]} \BibitemShut {NoStop}%
\bibitem [{\citenamefont {Fang}\ \emph {et~al.}(2020)\citenamefont {Fang}, \citenamefont {Metzger}, \citenamefont {Vurm}, \citenamefont {Aydi},\ and\ \citenamefont {Chomiuk}}]{Fang:2020bkm}%
  \BibitemOpen
  \bibfield  {author} {\bibinfo {author} {\bibfnamefont {K.}~\bibnamefont {Fang}}, \bibinfo {author} {\bibfnamefont {B.~D.}\ \bibnamefont {Metzger}}, \bibinfo {author} {\bibfnamefont {I.}~\bibnamefont {Vurm}}, \bibinfo {author} {\bibfnamefont {E.}~\bibnamefont {Aydi}}, \ and\ \bibinfo {author} {\bibfnamefont {L.}~\bibnamefont {Chomiuk}},\ }\href {\doibase 10.3847/1538-4357/abbc6e} {\bibfield  {journal} {\bibinfo  {journal} {Astrophys. J.}\ }\textbf {\bibinfo {volume} {904}},\ \bibinfo {pages} {4} (\bibinfo {year} {2020})},\ \Eprint {http://arxiv.org/abs/2007.15742} {arXiv:2007.15742 [astro-ph.HE]} \BibitemShut {NoStop}%
\bibitem [{\citenamefont {Liu}\ \emph {et~al.}(2020)\citenamefont {Liu}, \citenamefont {Xi},\ and\ \citenamefont {Wang}}]{Liu:2020isi}%
  \BibitemOpen
  \bibfield  {author} {\bibinfo {author} {\bibfnamefont {R.-Y.}\ \bibnamefont {Liu}}, \bibinfo {author} {\bibfnamefont {S.-Q.}\ \bibnamefont {Xi}}, \ and\ \bibinfo {author} {\bibfnamefont {X.-Y.}\ \bibnamefont {Wang}},\ }\href {\doibase 10.1103/PhysRevD.102.083028} {\bibfield  {journal} {\bibinfo  {journal} {Phys. Rev. D}\ }\textbf {\bibinfo {volume} {102}},\ \bibinfo {pages} {083028} (\bibinfo {year} {2020})},\ \Eprint {http://arxiv.org/abs/2011.03773} {arXiv:2011.03773 [astro-ph.HE]} \BibitemShut {NoStop}%
\bibitem [{\citenamefont {Winter}\ and\ \citenamefont {Lunardini}(2021)}]{Winter:2020ptf}%
  \BibitemOpen
  \bibfield  {author} {\bibinfo {author} {\bibfnamefont {W.}~\bibnamefont {Winter}}\ and\ \bibinfo {author} {\bibfnamefont {C.}~\bibnamefont {Lunardini}},\ }\href {\doibase 10.1038/s41550-021-01343-x} {\bibfield  {journal} {\bibinfo  {journal} {Nature Astron.}\ }\textbf {\bibinfo {volume} {5}},\ \bibinfo {pages} {472} (\bibinfo {year} {2021})},\ \Eprint {http://arxiv.org/abs/2005.06097} {arXiv:2005.06097 [astro-ph.HE]} \BibitemShut {NoStop}%
\bibitem [{\citenamefont {Wu}\ \emph {et~al.}(2022)\citenamefont {Wu}, \citenamefont {Mou}, \citenamefont {Wang}, \citenamefont {Wang},\ and\ \citenamefont {Li}}]{Wu:2021vaw}%
  \BibitemOpen
  \bibfield  {author} {\bibinfo {author} {\bibfnamefont {H.-J.}\ \bibnamefont {Wu}}, \bibinfo {author} {\bibfnamefont {G.}~\bibnamefont {Mou}}, \bibinfo {author} {\bibfnamefont {K.}~\bibnamefont {Wang}}, \bibinfo {author} {\bibfnamefont {W.}~\bibnamefont {Wang}}, \ and\ \bibinfo {author} {\bibfnamefont {Z.}~\bibnamefont {Li}},\ }\href {\doibase 10.1093/mnras/stac1621} {\bibfield  {journal} {\bibinfo  {journal} {Mon. Not. Roy. Astron. Soc.}\ }\textbf {\bibinfo {volume} {514}},\ \bibinfo {pages} {4406} (\bibinfo {year} {2022})},\ \Eprint {http://arxiv.org/abs/2112.01748} {arXiv:2112.01748 [astro-ph.HE]} \BibitemShut {NoStop}%
\bibitem [{\citenamefont {van Velzen}\ \emph {et~al.}(2021{\natexlab{b}})\citenamefont {van Velzen}, \citenamefont {Gezari}, \citenamefont {Hammerstein}, \citenamefont {Roth}, \citenamefont {Frederick}, \citenamefont {Ward}, \citenamefont {Hung}, \citenamefont {Cenko}, \citenamefont {Stein}, \citenamefont {Perley}, \citenamefont {Taggart}, \citenamefont {Foley}, \citenamefont {Sollerman}, \citenamefont {Blagorodnova}, \citenamefont {Andreoni}, \citenamefont {Bellm}, \citenamefont {Brinnel}, \citenamefont {De}, \citenamefont {Dekany}, \citenamefont {Feeney}, \citenamefont {Fremling}, \citenamefont {Giomi}, \citenamefont {Golkhou}, \citenamefont {Graham}, \citenamefont {Ho}, \citenamefont {Kasliwal}, \citenamefont {Kilpatrick}, \citenamefont {Kulkarni}, \citenamefont {Kupfer}, \citenamefont {Laher}, \citenamefont {Mahabal}, \citenamefont {Masci}, \citenamefont {Miller}, \citenamefont {Nordin}, \citenamefont {Riddle}, \citenamefont {Rusholme}, \citenamefont {van Santen}, \citenamefont {Sharma}, \citenamefont
  {Shupe},\ and\ \citenamefont {Soumagnac}}]{vanVelzen2021}%
  \BibitemOpen
  \bibfield  {author} {\bibinfo {author} {\bibfnamefont {S.}~\bibnamefont {van Velzen}}, \bibinfo {author} {\bibfnamefont {S.}~\bibnamefont {Gezari}}, \bibinfo {author} {\bibfnamefont {E.}~\bibnamefont {Hammerstein}}, \bibinfo {author} {\bibfnamefont {N.}~\bibnamefont {Roth}}, \bibinfo {author} {\bibfnamefont {S.}~\bibnamefont {Frederick}}, \bibinfo {author} {\bibfnamefont {C.}~\bibnamefont {Ward}}, \bibinfo {author} {\bibfnamefont {T.}~\bibnamefont {Hung}}, \bibinfo {author} {\bibfnamefont {S.~B.}\ \bibnamefont {Cenko}}, \bibinfo {author} {\bibfnamefont {R.}~\bibnamefont {Stein}}, \bibinfo {author} {\bibfnamefont {D.~A.}\ \bibnamefont {Perley}}, \bibinfo {author} {\bibfnamefont {K.}~\bibnamefont {Taggart}}, \bibinfo {author} {\bibfnamefont {R.~J.}\ \bibnamefont {Foley}}, \bibinfo {author} {\bibfnamefont {J.}~\bibnamefont {Sollerman}}, \bibinfo {author} {\bibfnamefont {N.}~\bibnamefont {Blagorodnova}}, \bibinfo {author} {\bibfnamefont {I.}~\bibnamefont {Andreoni}}, \bibinfo {author} {\bibfnamefont {E.~C.}\
  \bibnamefont {Bellm}}, \bibinfo {author} {\bibfnamefont {V.}~\bibnamefont {Brinnel}}, \bibinfo {author} {\bibfnamefont {K.}~\bibnamefont {De}}, \bibinfo {author} {\bibfnamefont {R.}~\bibnamefont {Dekany}}, \bibinfo {author} {\bibfnamefont {M.}~\bibnamefont {Feeney}}, \bibinfo {author} {\bibfnamefont {C.}~\bibnamefont {Fremling}}, \bibinfo {author} {\bibfnamefont {M.}~\bibnamefont {Giomi}}, \bibinfo {author} {\bibfnamefont {V.~Z.}\ \bibnamefont {Golkhou}}, \bibinfo {author} {\bibfnamefont {M.~J.}\ \bibnamefont {Graham}}, \bibinfo {author} {\bibfnamefont {A.~Y.~Q.}\ \bibnamefont {Ho}}, \bibinfo {author} {\bibfnamefont {M.~M.}\ \bibnamefont {Kasliwal}}, \bibinfo {author} {\bibfnamefont {C.~D.}\ \bibnamefont {Kilpatrick}}, \bibinfo {author} {\bibfnamefont {S.~R.}\ \bibnamefont {Kulkarni}}, \bibinfo {author} {\bibfnamefont {T.}~\bibnamefont {Kupfer}}, \bibinfo {author} {\bibfnamefont {R.~R.}\ \bibnamefont {Laher}}, \bibinfo {author} {\bibfnamefont {A.}~\bibnamefont {Mahabal}}, \bibinfo {author} {\bibfnamefont
  {F.~J.}\ \bibnamefont {Masci}}, \bibinfo {author} {\bibfnamefont {A.~A.}\ \bibnamefont {Miller}}, \bibinfo {author} {\bibfnamefont {J.}~\bibnamefont {Nordin}}, \bibinfo {author} {\bibfnamefont {R.}~\bibnamefont {Riddle}}, \bibinfo {author} {\bibfnamefont {B.}~\bibnamefont {Rusholme}}, \bibinfo {author} {\bibfnamefont {J.}~\bibnamefont {van Santen}}, \bibinfo {author} {\bibfnamefont {Y.}~\bibnamefont {Sharma}}, \bibinfo {author} {\bibfnamefont {D.~L.}\ \bibnamefont {Shupe}}, \ and\ \bibinfo {author} {\bibfnamefont {M.~T.}\ \bibnamefont {Soumagnac}},\ }\href {\doibase 10.3847/1538-4357/abc258} {\bibfield  {journal} {\bibinfo  {journal} {The Astrophysical Journal}\ }\textbf {\bibinfo {volume} {908}},\ \bibinfo {pages} {4} (\bibinfo {year} {2021}{\natexlab{b}})}\BibitemShut {NoStop}%
\bibitem [{\citenamefont {Reusch}\ \emph {et~al.}(2022{\natexlab{b}})\citenamefont {Reusch}, \citenamefont {Stein}, \citenamefont {Kowalski}, \citenamefont {van Velzen}, \citenamefont {Franckowiak}, \citenamefont {Lunardini}, \citenamefont {Murase}, \citenamefont {Winter}, \citenamefont {Miller-Jones}, \citenamefont {Kasliwal}, \citenamefont {Gilfanov}, \citenamefont {Garrappa}, \citenamefont {Paliya}, \citenamefont {Ahumada}, \citenamefont {Anand}, \citenamefont {Barbarino}, \citenamefont {Bellm}, \citenamefont {Brinnel}, \citenamefont {Buson}, \citenamefont {Cenko}, \citenamefont {Coughlin}, \citenamefont {De}, \citenamefont {Dekany}, \citenamefont {Frederick}, \citenamefont {Gal-Yam}, \citenamefont {Gezari}, \citenamefont {Giroletti}, \citenamefont {Graham}, \citenamefont {Karambelkar}, \citenamefont {Kimura}, \citenamefont {Kong}, \citenamefont {Kool}, \citenamefont {Laher}, \citenamefont {Medvedev}, \citenamefont {Necker}, \citenamefont {Nordin}, \citenamefont {Perley}, \citenamefont {Rigault},
  \citenamefont {Rusholme}, \citenamefont {Schulze}, \citenamefont {Schweyer}, \citenamefont {Singer}, \citenamefont {Sollerman}, \citenamefont {Strotjohann}, \citenamefont {Sunyaev}, \citenamefont {van Santen}, \citenamefont {Walters}, \citenamefont {Zhang},\ and\ \citenamefont {Zimmerman}}]{PhysRevLett.128.221101}%
  \BibitemOpen
  \bibfield  {author} {\bibinfo {author} {\bibfnamefont {S.}~\bibnamefont {Reusch}}, \bibinfo {author} {\bibfnamefont {R.}~\bibnamefont {Stein}}, \bibinfo {author} {\bibfnamefont {M.}~\bibnamefont {Kowalski}}, \bibinfo {author} {\bibfnamefont {S.}~\bibnamefont {van Velzen}}, \bibinfo {author} {\bibfnamefont {A.}~\bibnamefont {Franckowiak}}, \bibinfo {author} {\bibfnamefont {C.}~\bibnamefont {Lunardini}}, \bibinfo {author} {\bibfnamefont {K.}~\bibnamefont {Murase}}, \bibinfo {author} {\bibfnamefont {W.}~\bibnamefont {Winter}}, \bibinfo {author} {\bibfnamefont {J.~C.~A.}\ \bibnamefont {Miller-Jones}}, \bibinfo {author} {\bibfnamefont {M.~M.}\ \bibnamefont {Kasliwal}}, \bibinfo {author} {\bibfnamefont {M.}~\bibnamefont {Gilfanov}}, \bibinfo {author} {\bibfnamefont {S.}~\bibnamefont {Garrappa}}, \bibinfo {author} {\bibfnamefont {V.~S.}\ \bibnamefont {Paliya}}, \bibinfo {author} {\bibfnamefont {T.}~\bibnamefont {Ahumada}}, \bibinfo {author} {\bibfnamefont {S.}~\bibnamefont {Anand}}, \bibinfo {author}
  {\bibfnamefont {C.}~\bibnamefont {Barbarino}}, \bibinfo {author} {\bibfnamefont {E.~C.}\ \bibnamefont {Bellm}}, \bibinfo {author} {\bibfnamefont {V.}~\bibnamefont {Brinnel}}, \bibinfo {author} {\bibfnamefont {S.}~\bibnamefont {Buson}}, \bibinfo {author} {\bibfnamefont {S.~B.}\ \bibnamefont {Cenko}}, \bibinfo {author} {\bibfnamefont {M.~W.}\ \bibnamefont {Coughlin}}, \bibinfo {author} {\bibfnamefont {K.}~\bibnamefont {De}}, \bibinfo {author} {\bibfnamefont {R.}~\bibnamefont {Dekany}}, \bibinfo {author} {\bibfnamefont {S.}~\bibnamefont {Frederick}}, \bibinfo {author} {\bibfnamefont {A.}~\bibnamefont {Gal-Yam}}, \bibinfo {author} {\bibfnamefont {S.}~\bibnamefont {Gezari}}, \bibinfo {author} {\bibfnamefont {M.}~\bibnamefont {Giroletti}}, \bibinfo {author} {\bibfnamefont {M.~J.}\ \bibnamefont {Graham}}, \bibinfo {author} {\bibfnamefont {V.}~\bibnamefont {Karambelkar}}, \bibinfo {author} {\bibfnamefont {S.~S.}\ \bibnamefont {Kimura}}, \bibinfo {author} {\bibfnamefont {A.~K.~H.}\ \bibnamefont {Kong}}, \bibinfo
  {author} {\bibfnamefont {E.~C.}\ \bibnamefont {Kool}}, \bibinfo {author} {\bibfnamefont {R.~R.}\ \bibnamefont {Laher}}, \bibinfo {author} {\bibfnamefont {P.}~\bibnamefont {Medvedev}}, \bibinfo {author} {\bibfnamefont {J.}~\bibnamefont {Necker}}, \bibinfo {author} {\bibfnamefont {J.}~\bibnamefont {Nordin}}, \bibinfo {author} {\bibfnamefont {D.~A.}\ \bibnamefont {Perley}}, \bibinfo {author} {\bibfnamefont {M.}~\bibnamefont {Rigault}}, \bibinfo {author} {\bibfnamefont {B.}~\bibnamefont {Rusholme}}, \bibinfo {author} {\bibfnamefont {S.}~\bibnamefont {Schulze}}, \bibinfo {author} {\bibfnamefont {T.}~\bibnamefont {Schweyer}}, \bibinfo {author} {\bibfnamefont {L.~P.}\ \bibnamefont {Singer}}, \bibinfo {author} {\bibfnamefont {J.}~\bibnamefont {Sollerman}}, \bibinfo {author} {\bibfnamefont {N.~L.}\ \bibnamefont {Strotjohann}}, \bibinfo {author} {\bibfnamefont {R.}~\bibnamefont {Sunyaev}}, \bibinfo {author} {\bibfnamefont {J.}~\bibnamefont {van Santen}}, \bibinfo {author} {\bibfnamefont {R.}~\bibnamefont {Walters}},
  \bibinfo {author} {\bibfnamefont {B.~T.}\ \bibnamefont {Zhang}}, \ and\ \bibinfo {author} {\bibfnamefont {E.}~\bibnamefont {Zimmerman}},\ }\href {\doibase 10.1103/PhysRevLett.128.221101} {\bibfield  {journal} {\bibinfo  {journal} {Phys. Rev. Lett.}\ }\textbf {\bibinfo {volume} {128}},\ \bibinfo {pages} {221101} (\bibinfo {year} {2022}{\natexlab{b}})}\BibitemShut {NoStop}%
\bibitem [{\citenamefont {{Peebles}}(1972)}]{1972GReGr...3...63P}%
  \BibitemOpen
  \bibfield  {author} {\bibinfo {author} {\bibfnamefont {P.~J.~E.}\ \bibnamefont {{Peebles}}},\ }\href {\doibase 10.1007/BF00755923} {\bibfield  {journal} {\bibinfo  {journal} {General Relativity and Gravitation}\ }\textbf {\bibinfo {volume} {3}},\ \bibinfo {pages} {63} (\bibinfo {year} {1972})}\BibitemShut {NoStop}%
\bibitem [{\citenamefont {Quinlan}\ \emph {et~al.}(1995)\citenamefont {Quinlan}, \citenamefont {Hernquist},\ and\ \citenamefont {Sigurdsson}}]{Quinlan:1994ed}%
  \BibitemOpen
  \bibfield  {author} {\bibinfo {author} {\bibfnamefont {G.~D.}\ \bibnamefont {Quinlan}}, \bibinfo {author} {\bibfnamefont {L.}~\bibnamefont {Hernquist}}, \ and\ \bibinfo {author} {\bibfnamefont {S.}~\bibnamefont {Sigurdsson}},\ }\href {\doibase 10.1086/175295} {\bibfield  {journal} {\bibinfo  {journal} {Astrophys. J.}\ }\textbf {\bibinfo {volume} {440}},\ \bibinfo {pages} {554} (\bibinfo {year} {1995})},\ \Eprint {http://arxiv.org/abs/astro-ph/9407005} {arXiv:astro-ph/9407005} \BibitemShut {NoStop}%
\bibitem [{\citenamefont {Ullio}\ \emph {et~al.}(2001)\citenamefont {Ullio}, \citenamefont {Zhao},\ and\ \citenamefont {Kamionkowski}}]{Ullio:2001fb}%
  \BibitemOpen
  \bibfield  {author} {\bibinfo {author} {\bibfnamefont {P.}~\bibnamefont {Ullio}}, \bibinfo {author} {\bibfnamefont {H.}~\bibnamefont {Zhao}}, \ and\ \bibinfo {author} {\bibfnamefont {M.}~\bibnamefont {Kamionkowski}},\ }\href {\doibase 10.1103/PhysRevD.64.043504} {\bibfield  {journal} {\bibinfo  {journal} {Phys. Rev. D}\ }\textbf {\bibinfo {volume} {64}},\ \bibinfo {pages} {043504} (\bibinfo {year} {2001})},\ \Eprint {http://arxiv.org/abs/astro-ph/0101481} {arXiv:astro-ph/0101481} \BibitemShut {NoStop}%
\bibitem [{\citenamefont {Sadeghian}\ \emph {et~al.}(2013)\citenamefont {Sadeghian}, \citenamefont {Ferrer},\ and\ \citenamefont {Will}}]{Sadeghian_2013}%
  \BibitemOpen
  \bibfield  {author} {\bibinfo {author} {\bibfnamefont {L.}~\bibnamefont {Sadeghian}}, \bibinfo {author} {\bibfnamefont {F.}~\bibnamefont {Ferrer}}, \ and\ \bibinfo {author} {\bibfnamefont {C.~M.}\ \bibnamefont {Will}},\ }\href {\doibase 10.1103/physrevd.88.063522} {\bibfield  {journal} {\bibinfo  {journal} {Physical Review D}\ }\textbf {\bibinfo {volume} {88}} (\bibinfo {year} {2013}),\ 10.1103/physrevd.88.063522}\BibitemShut {NoStop}%
\bibitem [{\citenamefont {Ferrer}\ \emph {et~al.}(2017)\citenamefont {Ferrer}, \citenamefont {da~Rosa},\ and\ \citenamefont {Will}}]{Ferrer:2017xwm}%
  \BibitemOpen
  \bibfield  {author} {\bibinfo {author} {\bibfnamefont {F.}~\bibnamefont {Ferrer}}, \bibinfo {author} {\bibfnamefont {A.~M.}\ \bibnamefont {da~Rosa}}, \ and\ \bibinfo {author} {\bibfnamefont {C.~M.}\ \bibnamefont {Will}},\ }\href {\doibase 10.1103/PhysRevD.96.083014} {\bibfield  {journal} {\bibinfo  {journal} {Phys. Rev. D}\ }\textbf {\bibinfo {volume} {96}},\ \bibinfo {pages} {083014} (\bibinfo {year} {2017})},\ \Eprint {http://arxiv.org/abs/1707.06302} {arXiv:1707.06302 [astro-ph.CO]} \BibitemShut {NoStop}%
\bibitem [{\citenamefont {Boehm}\ \emph {et~al.}(2014)\citenamefont {Boehm}, \citenamefont {Schewtschenko}, \citenamefont {Wilkinson}, \citenamefont {Baugh},\ and\ \citenamefont {Pascoli}}]{Boehm:2014vja}%
  \BibitemOpen
  \bibfield  {author} {\bibinfo {author} {\bibfnamefont {C.}~\bibnamefont {Boehm}}, \bibinfo {author} {\bibfnamefont {J.~A.}\ \bibnamefont {Schewtschenko}}, \bibinfo {author} {\bibfnamefont {R.~J.}\ \bibnamefont {Wilkinson}}, \bibinfo {author} {\bibfnamefont {C.~M.}\ \bibnamefont {Baugh}}, \ and\ \bibinfo {author} {\bibfnamefont {S.}~\bibnamefont {Pascoli}},\ }\href {\doibase 10.1093/mnrasl/slu115} {\bibfield  {journal} {\bibinfo  {journal} {Mon. Not. Roy. Astron. Soc.}\ }\textbf {\bibinfo {volume} {445}},\ \bibinfo {pages} {L31} (\bibinfo {year} {2014})},\ \Eprint {http://arxiv.org/abs/1404.7012} {arXiv:1404.7012 [astro-ph.CO]} \BibitemShut {NoStop}%
\bibitem [{\citenamefont {Navarro}\ \emph {et~al.}(1997)\citenamefont {Navarro}, \citenamefont {Frenk},\ and\ \citenamefont {White}}]{Navarro:1996gj}%
  \BibitemOpen
  \bibfield  {author} {\bibinfo {author} {\bibfnamefont {J.~F.}\ \bibnamefont {Navarro}}, \bibinfo {author} {\bibfnamefont {C.~S.}\ \bibnamefont {Frenk}}, \ and\ \bibinfo {author} {\bibfnamefont {S.~D.~M.}\ \bibnamefont {White}},\ }\href {\doibase 10.1086/304888} {\bibfield  {journal} {\bibinfo  {journal} {Astrophys. J.}\ }\textbf {\bibinfo {volume} {490}},\ \bibinfo {pages} {493} (\bibinfo {year} {1997})},\ \Eprint {http://arxiv.org/abs/astro-ph/9611107} {arXiv:astro-ph/9611107} \BibitemShut {NoStop}%
\bibitem [{\citenamefont {Navarro}\ \emph {et~al.}(1996)\citenamefont {Navarro}, \citenamefont {Frenk},\ and\ \citenamefont {White}}]{Navarro:1995iw}%
  \BibitemOpen
  \bibfield  {author} {\bibinfo {author} {\bibfnamefont {J.~F.}\ \bibnamefont {Navarro}}, \bibinfo {author} {\bibfnamefont {C.~S.}\ \bibnamefont {Frenk}}, \ and\ \bibinfo {author} {\bibfnamefont {S.~D.~M.}\ \bibnamefont {White}},\ }\href {\doibase 10.1086/177173} {\bibfield  {journal} {\bibinfo  {journal} {Astrophys. J.}\ }\textbf {\bibinfo {volume} {462}},\ \bibinfo {pages} {563} (\bibinfo {year} {1996})},\ \Eprint {http://arxiv.org/abs/astro-ph/9508025} {arXiv:astro-ph/9508025} \BibitemShut {NoStop}%
\bibitem [{\citenamefont {Gorchtein}\ \emph {et~al.}(2010)\citenamefont {Gorchtein}, \citenamefont {Profumo},\ and\ \citenamefont {Ubaldi}}]{Gorchtein_2010}%
  \BibitemOpen
  \bibfield  {author} {\bibinfo {author} {\bibfnamefont {M.}~\bibnamefont {Gorchtein}}, \bibinfo {author} {\bibfnamefont {S.}~\bibnamefont {Profumo}}, \ and\ \bibinfo {author} {\bibfnamefont {L.}~\bibnamefont {Ubaldi}},\ }\href {\doibase 10.1103/physrevd.82.083514} {\bibfield  {journal} {\bibinfo  {journal} {Physical Review D}\ }\textbf {\bibinfo {volume} {82}} (\bibinfo {year} {2010}),\ 10.1103/physrevd.82.083514}\BibitemShut {NoStop}%
\bibitem [{\citenamefont {Lacroix}\ \emph {et~al.}(2017)\citenamefont {Lacroix}, \citenamefont {Karami}, \citenamefont {Broderick}, \citenamefont {Silk},\ and\ \citenamefont {B{\oe}hm}}]{Lacroix_2017}%
  \BibitemOpen
  \bibfield  {author} {\bibinfo {author} {\bibfnamefont {T.}~\bibnamefont {Lacroix}}, \bibinfo {author} {\bibfnamefont {M.}~\bibnamefont {Karami}}, \bibinfo {author} {\bibfnamefont {A.~E.}\ \bibnamefont {Broderick}}, \bibinfo {author} {\bibfnamefont {J.}~\bibnamefont {Silk}}, \ and\ \bibinfo {author} {\bibfnamefont {C.}~\bibnamefont {B{\oe}hm}},\ }\href {\doibase 10.1103/physrevd.96.063008} {\bibfield  {journal} {\bibinfo  {journal} {Physical Review D}\ }\textbf {\bibinfo {volume} {96}} (\bibinfo {year} {2017}),\ 10.1103/physrevd.96.063008}\BibitemShut {NoStop}%
\bibitem [{\citenamefont {Piana}\ \emph {et~al.}(2020)\citenamefont {Piana}, \citenamefont {Dayal}, \citenamefont {Volonteri},\ and\ \citenamefont {Choudhury}}]{Piana_2020}%
  \BibitemOpen
  \bibfield  {author} {\bibinfo {author} {\bibfnamefont {O.}~\bibnamefont {Piana}}, \bibinfo {author} {\bibfnamefont {P.}~\bibnamefont {Dayal}}, \bibinfo {author} {\bibfnamefont {M.}~\bibnamefont {Volonteri}}, \ and\ \bibinfo {author} {\bibfnamefont {T.~R.}\ \bibnamefont {Choudhury}},\ }\href {\doibase 10.1093/mnras/staa3363} {\bibfield  {journal} {\bibinfo  {journal} {Monthly Notices of the Royal Astronomical Society}\ }\textbf {\bibinfo {volume} {500}},\ \bibinfo {pages} {2146–2158} (\bibinfo {year} {2020})}\BibitemShut {NoStop}%
\bibitem [{\citenamefont {Lacroix}\ \emph {et~al.}(2015)\citenamefont {Lacroix}, \citenamefont {B{\oe}hm},\ and\ \citenamefont {Silk}}]{Lacroix_2015}%
  \BibitemOpen
  \bibfield  {author} {\bibinfo {author} {\bibfnamefont {T.}~\bibnamefont {Lacroix}}, \bibinfo {author} {\bibfnamefont {C.}~\bibnamefont {B{\oe}hm}}, \ and\ \bibinfo {author} {\bibfnamefont {J.}~\bibnamefont {Silk}},\ }\href {\doibase 10.1103/physrevd.92.043510} {\bibfield  {journal} {\bibinfo  {journal} {Physical Review D}\ }\textbf {\bibinfo {volume} {92}} (\bibinfo {year} {2015}),\ 10.1103/physrevd.92.043510}\BibitemShut {NoStop}%
\bibitem [{\citenamefont {Berryman}\ \emph {et~al.}(2023)\citenamefont {Berryman} \emph {et~al.}}]{Berryman:2022hds}%
  \BibitemOpen
  \bibfield  {author} {\bibinfo {author} {\bibfnamefont {J.~M.}\ \bibnamefont {Berryman}} \emph {et~al.},\ }\href {\doibase 10.1016/j.dark.2023.101267} {\bibfield  {journal} {\bibinfo  {journal} {Phys. Dark Univ.}\ }\textbf {\bibinfo {volume} {42}},\ \bibinfo {pages} {101267} (\bibinfo {year} {2023})},\ \Eprint {http://arxiv.org/abs/2203.01955} {arXiv:2203.01955 [hep-ph]} \BibitemShut {NoStop}%
\bibitem [{\citenamefont {Berlin}\ \emph {et~al.}(2019)\citenamefont {Berlin}, \citenamefont {Blinov}, \citenamefont {Krnjaic}, \citenamefont {Schuster},\ and\ \citenamefont {Toro}}]{Berlin:2018bsc}%
  \BibitemOpen
  \bibfield  {author} {\bibinfo {author} {\bibfnamefont {A.}~\bibnamefont {Berlin}}, \bibinfo {author} {\bibfnamefont {N.}~\bibnamefont {Blinov}}, \bibinfo {author} {\bibfnamefont {G.}~\bibnamefont {Krnjaic}}, \bibinfo {author} {\bibfnamefont {P.}~\bibnamefont {Schuster}}, \ and\ \bibinfo {author} {\bibfnamefont {N.}~\bibnamefont {Toro}},\ }\href {\doibase 10.1103/PhysRevD.99.075001} {\bibfield  {journal} {\bibinfo  {journal} {Phys. Rev. D}\ }\textbf {\bibinfo {volume} {99}},\ \bibinfo {pages} {075001} (\bibinfo {year} {2019})},\ \Eprint {http://arxiv.org/abs/1807.01730} {arXiv:1807.01730 [hep-ph]} \BibitemShut {NoStop}%
\bibitem [{\citenamefont {Berryman}\ \emph {et~al.}(2018)\citenamefont {Berryman}, \citenamefont {De~Gouv\^ea}, \citenamefont {Kelly},\ and\ \citenamefont {Zhang}}]{Berryman:2018ogk}%
  \BibitemOpen
  \bibfield  {author} {\bibinfo {author} {\bibfnamefont {J.~M.}\ \bibnamefont {Berryman}}, \bibinfo {author} {\bibfnamefont {A.}~\bibnamefont {De~Gouv\^ea}}, \bibinfo {author} {\bibfnamefont {K.~J.}\ \bibnamefont {Kelly}}, \ and\ \bibinfo {author} {\bibfnamefont {Y.}~\bibnamefont {Zhang}},\ }\href {\doibase 10.1103/PhysRevD.97.075030} {\bibfield  {journal} {\bibinfo  {journal} {Phys. Rev. D}\ }\textbf {\bibinfo {volume} {97}},\ \bibinfo {pages} {075030} (\bibinfo {year} {2018})},\ \Eprint {http://arxiv.org/abs/1802.00009} {arXiv:1802.00009 [hep-ph]} \BibitemShut {NoStop}%
\bibitem [{\citenamefont {de~Gouv\^ea}\ \emph {et~al.}(2020)\citenamefont {de~Gouv\^ea}, \citenamefont {Dev}, \citenamefont {Dutta}, \citenamefont {Ghosh}, \citenamefont {Han},\ and\ \citenamefont {Zhang}}]{deGouvea:2019qaz}%
  \BibitemOpen
  \bibfield  {author} {\bibinfo {author} {\bibfnamefont {A.}~\bibnamefont {de~Gouv\^ea}}, \bibinfo {author} {\bibfnamefont {P.~S.~B.}\ \bibnamefont {Dev}}, \bibinfo {author} {\bibfnamefont {B.}~\bibnamefont {Dutta}}, \bibinfo {author} {\bibfnamefont {T.}~\bibnamefont {Ghosh}}, \bibinfo {author} {\bibfnamefont {T.}~\bibnamefont {Han}}, \ and\ \bibinfo {author} {\bibfnamefont {Y.}~\bibnamefont {Zhang}},\ }\href {\doibase 10.1007/JHEP07(2020)142} {\bibfield  {journal} {\bibinfo  {journal} {JHEP}\ }\textbf {\bibinfo {volume} {07}},\ \bibinfo {pages} {142} (\bibinfo {year} {2020})},\ \Eprint {http://arxiv.org/abs/1910.01132} {arXiv:1910.01132 [hep-ph]} \BibitemShut {NoStop}%
\bibitem [{\citenamefont {Laha}\ \emph {et~al.}(2014)\citenamefont {Laha}, \citenamefont {Dasgupta},\ and\ \citenamefont {Beacom}}]{Laha:2013xua}%
  \BibitemOpen
  \bibfield  {author} {\bibinfo {author} {\bibfnamefont {R.}~\bibnamefont {Laha}}, \bibinfo {author} {\bibfnamefont {B.}~\bibnamefont {Dasgupta}}, \ and\ \bibinfo {author} {\bibfnamefont {J.~F.}\ \bibnamefont {Beacom}},\ }\href {\doibase 10.1103/PhysRevD.89.093025} {\bibfield  {journal} {\bibinfo  {journal} {Phys. Rev. D}\ }\textbf {\bibinfo {volume} {89}},\ \bibinfo {pages} {093025} (\bibinfo {year} {2014})},\ \Eprint {http://arxiv.org/abs/1304.3460} {arXiv:1304.3460 [hep-ph]} \BibitemShut {NoStop}%
\bibitem [{\citenamefont {Ibe}\ \emph {et~al.}(2017)\citenamefont {Ibe}, \citenamefont {Nakano},\ and\ \citenamefont {Suzuki}}]{Ibe:2016dir}%
  \BibitemOpen
  \bibfield  {author} {\bibinfo {author} {\bibfnamefont {M.}~\bibnamefont {Ibe}}, \bibinfo {author} {\bibfnamefont {W.}~\bibnamefont {Nakano}}, \ and\ \bibinfo {author} {\bibfnamefont {M.}~\bibnamefont {Suzuki}},\ }\href {\doibase 10.1103/PhysRevD.95.055022} {\bibfield  {journal} {\bibinfo  {journal} {Phys. Rev. D}\ }\textbf {\bibinfo {volume} {95}},\ \bibinfo {pages} {055022} (\bibinfo {year} {2017})},\ \Eprint {http://arxiv.org/abs/1611.08460} {arXiv:1611.08460 [hep-ph]} \BibitemShut {NoStop}%
\bibitem [{\citenamefont {Bakhti}\ and\ \citenamefont {Farzan}(2017)}]{Bakhti:2017jhm}%
  \BibitemOpen
  \bibfield  {author} {\bibinfo {author} {\bibfnamefont {P.}~\bibnamefont {Bakhti}}\ and\ \bibinfo {author} {\bibfnamefont {Y.}~\bibnamefont {Farzan}},\ }\href {\doibase 10.1103/PhysRevD.95.095008} {\bibfield  {journal} {\bibinfo  {journal} {Phys. Rev. D}\ }\textbf {\bibinfo {volume} {95}},\ \bibinfo {pages} {095008} (\bibinfo {year} {2017})},\ \Eprint {http://arxiv.org/abs/1702.04187} {arXiv:1702.04187 [hep-ph]} \BibitemShut {NoStop}%
\bibitem [{\citenamefont {Burgess}\ and\ \citenamefont {Cline}(1993)}]{Burgess:1992dt}%
  \BibitemOpen
  \bibfield  {author} {\bibinfo {author} {\bibfnamefont {C.~P.}\ \bibnamefont {Burgess}}\ and\ \bibinfo {author} {\bibfnamefont {J.~M.}\ \bibnamefont {Cline}},\ }\href {\doibase 10.1016/0370-2693(93)91720-8} {\bibfield  {journal} {\bibinfo  {journal} {Phys. Lett. B}\ }\textbf {\bibinfo {volume} {298}},\ \bibinfo {pages} {141} (\bibinfo {year} {1993})},\ \Eprint {http://arxiv.org/abs/hep-ph/9209299} {arXiv:hep-ph/9209299} \BibitemShut {NoStop}%
\bibitem [{\citenamefont {Brune}\ and\ \citenamefont {P\"as}(2019)}]{Brune:2018sab}%
  \BibitemOpen
  \bibfield  {author} {\bibinfo {author} {\bibfnamefont {T.}~\bibnamefont {Brune}}\ and\ \bibinfo {author} {\bibfnamefont {H.}~\bibnamefont {P\"as}},\ }\href {\doibase 10.1103/PhysRevD.99.096005} {\bibfield  {journal} {\bibinfo  {journal} {Phys. Rev. D}\ }\textbf {\bibinfo {volume} {99}},\ \bibinfo {pages} {096005} (\bibinfo {year} {2019})},\ \Eprint {http://arxiv.org/abs/1808.08158} {arXiv:1808.08158 [hep-ph]} \BibitemShut {NoStop}%
\bibitem [{\citenamefont {Cepedello}\ \emph {et~al.}(2019)\citenamefont {Cepedello}, \citenamefont {Deppisch}, \citenamefont {Gonz\'alez}, \citenamefont {Hati},\ and\ \citenamefont {Hirsch}}]{Cepedello:2018zvr}%
  \BibitemOpen
  \bibfield  {author} {\bibinfo {author} {\bibfnamefont {R.}~\bibnamefont {Cepedello}}, \bibinfo {author} {\bibfnamefont {F.~F.}\ \bibnamefont {Deppisch}}, \bibinfo {author} {\bibfnamefont {L.}~\bibnamefont {Gonz\'alez}}, \bibinfo {author} {\bibfnamefont {C.}~\bibnamefont {Hati}}, \ and\ \bibinfo {author} {\bibfnamefont {M.}~\bibnamefont {Hirsch}},\ }\href {\doibase 10.1103/PhysRevLett.122.181801} {\bibfield  {journal} {\bibinfo  {journal} {Phys. Rev. Lett.}\ }\textbf {\bibinfo {volume} {122}},\ \bibinfo {pages} {181801} (\bibinfo {year} {2019})},\ \Eprint {http://arxiv.org/abs/1811.00031} {arXiv:1811.00031 [hep-ph]} \BibitemShut {NoStop}%
\bibitem [{\citenamefont {Huang}\ \emph {et~al.}(2018)\citenamefont {Huang}, \citenamefont {Ohlsson},\ and\ \citenamefont {Zhou}}]{Huang:2017egl}%
  \BibitemOpen
  \bibfield  {author} {\bibinfo {author} {\bibfnamefont {G.-y.}\ \bibnamefont {Huang}}, \bibinfo {author} {\bibfnamefont {T.}~\bibnamefont {Ohlsson}}, \ and\ \bibinfo {author} {\bibfnamefont {S.}~\bibnamefont {Zhou}},\ }\href {\doibase 10.1103/PhysRevD.97.075009} {\bibfield  {journal} {\bibinfo  {journal} {Phys. Rev. D}\ }\textbf {\bibinfo {volume} {97}},\ \bibinfo {pages} {075009} (\bibinfo {year} {2018})},\ \Eprint {http://arxiv.org/abs/1712.04792} {arXiv:1712.04792 [hep-ph]} \BibitemShut {NoStop}%
\bibitem [{\citenamefont {Grohs}\ \emph {et~al.}(2020)\citenamefont {Grohs}, \citenamefont {Fuller},\ and\ \citenamefont {Sen}}]{Grohs:2020xxd}%
  \BibitemOpen
  \bibfield  {author} {\bibinfo {author} {\bibfnamefont {E.}~\bibnamefont {Grohs}}, \bibinfo {author} {\bibfnamefont {G.~M.}\ \bibnamefont {Fuller}}, \ and\ \bibinfo {author} {\bibfnamefont {M.}~\bibnamefont {Sen}},\ }\href {\doibase 10.1088/1475-7516/2020/07/001} {\bibfield  {journal} {\bibinfo  {journal} {JCAP}\ }\textbf {\bibinfo {volume} {07}},\ \bibinfo {pages} {001} (\bibinfo {year} {2020})},\ \Eprint {http://arxiv.org/abs/2002.08557} {arXiv:2002.08557 [astro-ph.CO]} \BibitemShut {NoStop}%
\bibitem [{\citenamefont {Chauhan}\ \emph {et~al.}(2023)\citenamefont {Chauhan}, \citenamefont {Dev},\ and\ \citenamefont {Xu}}]{Chauhan:2022iuh}%
  \BibitemOpen
  \bibfield  {author} {\bibinfo {author} {\bibfnamefont {G.}~\bibnamefont {Chauhan}}, \bibinfo {author} {\bibfnamefont {P.~S.~B.}\ \bibnamefont {Dev}}, \ and\ \bibinfo {author} {\bibfnamefont {X.-J.}\ \bibnamefont {Xu}},\ }\href {\doibase 10.1016/j.physletb.2023.137907} {\bibfield  {journal} {\bibinfo  {journal} {Phys. Lett. B}\ }\textbf {\bibinfo {volume} {841}},\ \bibinfo {pages} {137907} (\bibinfo {year} {2023})},\ \Eprint {http://arxiv.org/abs/2204.11876} {arXiv:2204.11876 [hep-ph]} \BibitemShut {NoStop}%
\bibitem [{\citenamefont {Bertoni}\ \emph {et~al.}(2015)\citenamefont {Bertoni}, \citenamefont {Ipek}, \citenamefont {McKeen},\ and\ \citenamefont {Nelson}}]{Bertoni:2014mva}%
  \BibitemOpen
  \bibfield  {author} {\bibinfo {author} {\bibfnamefont {B.}~\bibnamefont {Bertoni}}, \bibinfo {author} {\bibfnamefont {S.}~\bibnamefont {Ipek}}, \bibinfo {author} {\bibfnamefont {D.}~\bibnamefont {McKeen}}, \ and\ \bibinfo {author} {\bibfnamefont {A.~E.}\ \bibnamefont {Nelson}},\ }\href {\doibase 10.1007/JHEP04(2015)170} {\bibfield  {journal} {\bibinfo  {journal} {JHEP}\ }\textbf {\bibinfo {volume} {04}},\ \bibinfo {pages} {170} (\bibinfo {year} {2015})},\ \Eprint {http://arxiv.org/abs/1412.3113} {arXiv:1412.3113 [hep-ph]} \BibitemShut {NoStop}%
\bibitem [{\citenamefont {Di~Valentino}\ \emph {et~al.}(2018)\citenamefont {Di~Valentino}, \citenamefont {B\o{}ehm}, \citenamefont {Hivon},\ and\ \citenamefont {Bouchet}}]{DiValentino:2017oaw}%
  \BibitemOpen
  \bibfield  {author} {\bibinfo {author} {\bibfnamefont {E.}~\bibnamefont {Di~Valentino}}, \bibinfo {author} {\bibfnamefont {C.}~\bibnamefont {B\o{}ehm}}, \bibinfo {author} {\bibfnamefont {E.}~\bibnamefont {Hivon}}, \ and\ \bibinfo {author} {\bibfnamefont {F.~R.}\ \bibnamefont {Bouchet}},\ }\href {\doibase 10.1103/PhysRevD.97.043513} {\bibfield  {journal} {\bibinfo  {journal} {Phys. Rev. D}\ }\textbf {\bibinfo {volume} {97}},\ \bibinfo {pages} {043513} (\bibinfo {year} {2018})},\ \Eprint {http://arxiv.org/abs/1710.02559} {arXiv:1710.02559 [astro-ph.CO]} \BibitemShut {NoStop}%
\bibitem [{\citenamefont {Olivares-Del~Campo}\ \emph {et~al.}(2018)\citenamefont {Olivares-Del~Campo}, \citenamefont {B\oe{}hm}, \citenamefont {Palomares-Ruiz},\ and\ \citenamefont {Pascoli}}]{Olivares-DelCampo:2017feq}%
  \BibitemOpen
  \bibfield  {author} {\bibinfo {author} {\bibfnamefont {A.}~\bibnamefont {Olivares-Del~Campo}}, \bibinfo {author} {\bibfnamefont {C.}~\bibnamefont {B\oe{}hm}}, \bibinfo {author} {\bibfnamefont {S.}~\bibnamefont {Palomares-Ruiz}}, \ and\ \bibinfo {author} {\bibfnamefont {S.}~\bibnamefont {Pascoli}},\ }\href {\doibase 10.1103/PhysRevD.97.075039} {\bibfield  {journal} {\bibinfo  {journal} {Phys. Rev. D}\ }\textbf {\bibinfo {volume} {97}},\ \bibinfo {pages} {075039} (\bibinfo {year} {2018})},\ \Eprint {http://arxiv.org/abs/1711.05283} {arXiv:1711.05283 [hep-ph]} \BibitemShut {NoStop}%
\bibitem [{\citenamefont {Hooper}\ and\ \citenamefont {Lucca}(2022)}]{Hooper:2021rjc}%
  \BibitemOpen
  \bibfield  {author} {\bibinfo {author} {\bibfnamefont {D.~C.}\ \bibnamefont {Hooper}}\ and\ \bibinfo {author} {\bibfnamefont {M.}~\bibnamefont {Lucca}},\ }\href {\doibase 10.1103/PhysRevD.105.103504} {\bibfield  {journal} {\bibinfo  {journal} {Phys. Rev. D}\ }\textbf {\bibinfo {volume} {105}},\ \bibinfo {pages} {103504} (\bibinfo {year} {2022})},\ \Eprint {http://arxiv.org/abs/2110.04024} {arXiv:2110.04024 [astro-ph.CO]} \BibitemShut {NoStop}%
\bibitem [{\citenamefont {Brax}\ \emph {et~al.}(2023{\natexlab{b}})\citenamefont {Brax}, \citenamefont {van~de Bruck}, \citenamefont {Di~Valentino}, \citenamefont {Giar\`e},\ and\ \citenamefont {Trojanowski}}]{Brax:2023rrf}%
  \BibitemOpen
  \bibfield  {author} {\bibinfo {author} {\bibfnamefont {P.}~\bibnamefont {Brax}}, \bibinfo {author} {\bibfnamefont {C.}~\bibnamefont {van~de Bruck}}, \bibinfo {author} {\bibfnamefont {E.}~\bibnamefont {Di~Valentino}}, \bibinfo {author} {\bibfnamefont {W.}~\bibnamefont {Giar\`e}}, \ and\ \bibinfo {author} {\bibfnamefont {S.}~\bibnamefont {Trojanowski}},\ }\href {\doibase 10.1093/mnrasl/slad157} {\bibfield  {journal} {\bibinfo  {journal} {Mon. Not. Roy. Astron. Soc.}\ }\textbf {\bibinfo {volume} {527}},\ \bibinfo {pages} {L122} (\bibinfo {year} {2023}{\natexlab{b}})},\ \Eprint {http://arxiv.org/abs/2303.16895} {arXiv:2303.16895 [astro-ph.CO]} \BibitemShut {NoStop}%
\bibitem [{\citenamefont {Akita}\ and\ \citenamefont {Ando}(2023)}]{Akita:2023yga}%
  \BibitemOpen
  \bibfield  {author} {\bibinfo {author} {\bibfnamefont {K.}~\bibnamefont {Akita}}\ and\ \bibinfo {author} {\bibfnamefont {S.}~\bibnamefont {Ando}},\ }\href {\doibase 10.1088/1475-7516/2023/11/037} {\bibfield  {journal} {\bibinfo  {journal} {JCAP}\ }\textbf {\bibinfo {volume} {11}},\ \bibinfo {pages} {037} (\bibinfo {year} {2023})},\ \Eprint {http://arxiv.org/abs/2305.01913} {arXiv:2305.01913 [astro-ph.CO]} \BibitemShut {NoStop}%
\bibitem [{\citenamefont {Giar\`e}\ \emph {et~al.}(2023)\citenamefont {Giar\`e}, \citenamefont {G\'omez-Valent}, \citenamefont {Di~Valentino},\ and\ \citenamefont {van~de Bruck}}]{Giare:2023qqn}%
  \BibitemOpen
  \bibfield  {author} {\bibinfo {author} {\bibfnamefont {W.}~\bibnamefont {Giar\`e}}, \bibinfo {author} {\bibfnamefont {A.}~\bibnamefont {G\'omez-Valent}}, \bibinfo {author} {\bibfnamefont {E.}~\bibnamefont {Di~Valentino}}, \ and\ \bibinfo {author} {\bibfnamefont {C.}~\bibnamefont {van~de Bruck}},\ }\href@noop {} {\  (\bibinfo {year} {2023})},\ \Eprint {http://arxiv.org/abs/2311.09116} {arXiv:2311.09116 [astro-ph.CO]} \BibitemShut {NoStop}%
\bibitem [{\citenamefont {Vogel}\ and\ \citenamefont {Redondo}(2014)}]{Vogel:2013raa}%
  \BibitemOpen
  \bibfield  {author} {\bibinfo {author} {\bibfnamefont {H.}~\bibnamefont {Vogel}}\ and\ \bibinfo {author} {\bibfnamefont {J.}~\bibnamefont {Redondo}},\ }\href {\doibase 10.1088/1475-7516/2014/02/029} {\bibfield  {journal} {\bibinfo  {journal} {JCAP}\ }\textbf {\bibinfo {volume} {02}},\ \bibinfo {pages} {029} (\bibinfo {year} {2014})},\ \Eprint {http://arxiv.org/abs/1311.2600} {arXiv:1311.2600 [hep-ph]} \BibitemShut {NoStop}%
\bibitem [{\citenamefont {Davidson}\ \emph {et~al.}(2000)\citenamefont {Davidson}, \citenamefont {Hannestad},\ and\ \citenamefont {Raffelt}}]{Davidson:2000hf}%
  \BibitemOpen
  \bibfield  {author} {\bibinfo {author} {\bibfnamefont {S.}~\bibnamefont {Davidson}}, \bibinfo {author} {\bibfnamefont {S.}~\bibnamefont {Hannestad}}, \ and\ \bibinfo {author} {\bibfnamefont {G.}~\bibnamefont {Raffelt}},\ }\href {\doibase 10.1088/1126-6708/2000/05/003} {\bibfield  {journal} {\bibinfo  {journal} {JHEP}\ }\textbf {\bibinfo {volume} {05}},\ \bibinfo {pages} {003} (\bibinfo {year} {2000})},\ \Eprint {http://arxiv.org/abs/hep-ph/0001179} {arXiv:hep-ph/0001179} \BibitemShut {NoStop}%
\bibitem [{\citenamefont {Liu}\ \emph {et~al.}(2019)\citenamefont {Liu}, \citenamefont {Outmezguine}, \citenamefont {Redigolo},\ and\ \citenamefont {Volansky}}]{Liu:2019knx}%
  \BibitemOpen
  \bibfield  {author} {\bibinfo {author} {\bibfnamefont {H.}~\bibnamefont {Liu}}, \bibinfo {author} {\bibfnamefont {N.~J.}\ \bibnamefont {Outmezguine}}, \bibinfo {author} {\bibfnamefont {D.}~\bibnamefont {Redigolo}}, \ and\ \bibinfo {author} {\bibfnamefont {T.}~\bibnamefont {Volansky}},\ }\href {\doibase 10.1103/PhysRevD.100.123011} {\bibfield  {journal} {\bibinfo  {journal} {Phys. Rev. D}\ }\textbf {\bibinfo {volume} {100}},\ \bibinfo {pages} {123011} (\bibinfo {year} {2019})},\ \Eprint {http://arxiv.org/abs/1908.06986} {arXiv:1908.06986 [hep-ph]} \BibitemShut {NoStop}%
\end{thebibliography}%

\end{document}